\documentclass{aa}
\usepackage{psfig}

\usepackage[dvips]{graphics}

\begin{document}
\title{Young massive stars in the ISOGAL survey
\thanks{This is paper no. 11  in a refereed
journal based on data from the ISOGAL project}\fnmsep
\thanks{Based on observations with ISO, an ESA project with instruments
funded by ESA Member States (especially the PI countries: France,
Germany, the Netherlands and the United Kingdom) and with the
participation of ISAS and NASA}\fnmsep
\thanks{Table~3 is only available in electronic form
at the CDS via anonymous ftp to cdsarc.u-strasbg.fr (130.79.128.5)
or via http://cdsweb.u-strasbg.fr/cgi-bin/qcat?J/A+A/}
}
\subtitle{II. The catalogue of bright YSO candidates}

\author{Marcello Felli\inst{1},
Leonardo Testi\inst{1}, Fr\'ed\'eric Schuller\inst{2}, Alain Omont\inst{2}}
\offprints{L. Testi, lt@arcetri.astro.it}

\institute{
Osservatorio Astrofisico di Arcetri, INAF, Largo E. Fermi, 5, 50125 Firenze, Italy
\and
Institut d'Astrophysique de Paris, CNRS, 98 bis Bd Arago, F-75014 Paris, France
}

\date{Received; Accepted }

\abstract{
The 7 and 15~$\mu$m observations of selected fields in the Galactic Plane
obtained with ISOCAM during  the ISOGAL program offer an unique possibility to
search for previously unknown YSOs,  undetected by IRAS because of lower
sensitivity or confusion problems.  In a previous paper (Felli et
al.~\cite{FCTOS00}) we established criteria of general validity to select YSOs
from the much larger population of Post Main Sequence (Post-MS) stars present
in the  ISOGAL fields by comparing radio and IR observations of five fields
located at l$\sim$+45$^{\circ}$.  The selection was based primarily on  the
position of the point sources in the [15] - [7]--[15] diagram, which involves
only ISOGAL data and allows to find possible YSOs using the survey data alone.
In the present work we revise the adopted criteria by comparing
radio-identified UC HII regions and ISOGAL observations over a much larger
region.  The main indications of the previous analysis are confirmed, but the
criteria for selecting YSO candidates had to be revised to select only bright
objects, in order to limit the contamination of the sample by Post-MS stars.
The revised criteria ([15]$\le$4.5, [7]--[15]$\ge$1.8) are then used to extract
YSO candidates from the ISOGAL Point Source Catalogue in preparation.
We select a total of 715 YSO candidates, corresponding to $\sim$2\% of the
sources with good detections at 7 and 15~$\mu$m.  The results are presented in
a table form that provides an unique input list of small diameter,
$\le 6$\arcsec, Galactic YSO
candidates. The  global properties of the sample of YSO candidates are briefly
discussed.
}

\titlerunning{Galactic YSOs}
\authorrunning{M. Felli et al.}
\maketitle

\section{Introduction}
The first steps of star formation are the pre-stellar and proto-stellar phases,
corresponding to the fragmentation and the gravitational collapse of a dense
core in a molecular cloud, before a formed star appears.  The IR spectrum of
pre- and proto-stellar cloud cores originates from a very cool dust envelope,
and  they
are observed in absorption against the diffuse background at 7 and 15~$\mu$m
(Molinari et al.~\cite{MTBCP}, Bacmann et al.~\cite{BAPABW}). Some of the 
infrared dark clouds detected by the ISOGAL and MSX surveys may indeed 
be the same type of objects (P\'erault et al.~\cite{Pea96}; Carey et
al.~\cite{Cea98}).

The next step is commonly referred to as Young Stellar Object (YSO), to
indicate the phase when the formed star is deeply embedded in a thick dusty
envelope, or when it is hidden by an optically thick disk, remnants of the
molecular cloud from which it was formed.  This envelope or disk absorbs all
the stellar radiation, making the YSO undetectable in the visible range, and
re-emits in the IR, thus making it shine as a bright IR source.

YSOs may have widely different luminosities and masses, ranging from a fraction
of a solar mass to 100 solar masses. The YSOs associated with the earliest
spectral type stars (earlier than B3) can be searched both in the radio
continuum, where the ionised gas of the Ultra Compact HII region (UC HII)
produces free-free emission, and in the Middle and Far IR (MIR and FIR,
respectively), where the dust emits.  Very bright UC HII regions have been
extensively studied (see e.g. Churchwell~\cite{C91} for a review) and models
have been developed to explain the Spectral Energy Distribution (SED) and the
spatial morphology of the emission at different wavelengths (Scoville \&
Kwan~\cite{SK76}, Rowan-Robinson~\cite{RR80}, Churchwell et al.~\cite{CWW90},
Ivezi\'c \& Elitzur~\cite{IE97}, Faison et al.~\cite{FCHHLR98}, Miroshnichenko
et al.~\cite{MIVE}, Feldt et al.~\cite{Fea99}).

YSOs associated with later spectral type stars (later than B3) can only be
detected in the FIR (and sub-mm) thanks to dust emission, as the radio emission
decreases sharply since the star doesn't supply enough Lyman continuum photons.
When the YSO becomes visible in the optical range, with a luminosity in the
range $\sim$10 to 10$^3$ L$_{\odot}$, it is called Herbig Ae/Be star. The IR
excess comes from a disk, an envelope or a combination of the two (see e.g.
Berrilli et al.~\cite{BCILNS92}, Hillenbrand et al.~\cite{HSVK92}, Pezzuto et
al.~\cite{PSL97}). At lower luminosity (of the order of one L$_{\odot}$) an
evolutionary track is now well established, going from Class I objects, with
considerable IR emission, to class II, when a T Tauri star becomes visible in
the optical range but is still surrounded by a disk, and finally to class III
when the stellar photosphere becomes visible (Andr\'e et al.~\cite{AWTB93};
Lada \& Wilking~\cite{LW84}; Lada~\cite{LADA87}; Lada~\cite{Lada99}
and Natta~\cite{Natta99} for recent review papers).

Finally, YSOs  with masses in the brown dwarf range have now been detected in
the MIR (Olofsson et al.~\cite{Olofsson99}; Persi et al.~\cite{Pea00};
Comer\'on et al.~\cite{CNK00}). The IR properties of these objects are similar
to those of more massive YSOs, suggesting that also in these cases the MIR
emission is produced by a circumstellar disk (Natta \& Testi~\cite{NT01},
Testi et al.~\cite{Tea02}).
Indeed, even though the spread of luminosities is very large, all these types
of YSOs show rather similar SEDs in the IR range, because both the emission
mechanism (re-radiation at lower temperature of the stellar emission absorbed
by the dust) and the geometry of the dust components are similar. Thus the
colours depend mainly on the overall optical depth in the dust but not on the
stellar luminosity (Ivezi\'c \& Elitzur~\cite{IE97}).
In particular, the systematic study of nearby star forming regions in the
frame of the ISOCAM guaranteed time program (Nordh et al.~\cite{Nordh96},
Bontemps et al.~\cite{Bont01} and references therein) have shown that,
especially for low mass YSOs, there is a clear cut between dusty YSOs of class
I and II and the young stars without much dust (class III) at
Log$_{10}$[S$_{15}$/S$_7$]~=~--0.2, i.e. [7]--[15]~=~1.1.

The ISOCAM observations of the Galactic Plane carried out during
the ISOGAL program offer an unique opportunity for an unbiased search
of YSOs. However, their identification requires to be able to
separate them from the much larger population of
Post-MS  stars in the Galactic Plane, which also have some IR excess due to
dust in a circumstellar envelope produced by mass loss,
but are in an entirely different evolutionary stage.

In a previous work (Felli et al.~\cite{FCTOS00}, hereafter Paper~II)
a comparison
between Very Large Array 3.6 and 6 cm radio continuum
observations (Testi et al.~\cite{TFT99}, hereafter Paper~I)
and ISOGAL observations of five
Galactic fields  at l $\sim$ +45$^{\circ}$,
in the two broad band filters
LW2 (5.5-8.5~$\mu$m) and LW3 (12-18~$\mu$m) was used
to establish general
criteria that allow the  identification of  YSOs.
These criteria were then used to extract the YSOs from the preliminary lists
of ISOGAL sources in those fields.

In the present paper we extend the comparison between UC HII regions
(or massive YSOs) identified in the radio continuum
and ISOGAL observations to a much larger region of the Galactic Plane
covered uniformly  with the Very Large Array at 6 cm by the BWHZ survey
(Becker et al.~\cite{bec94}). The much larger sample of radio-identified
YSOs allows
a better refinement of the identification criteria.
In Section~\ref{scat} the revised criteria are then used to extract the YSOs
from the ISOGAL Point Source Catalogue (Omont et al. in preparation, Schuller
et al. in preparation).
Finally, the galactic distribution and global properties of this sample
are briefly discussed.

\section{The ISOGAL Catalogue}
\label{sigcat}

\subsection{The ISOGAL Data}
\label{isog_data}
The ISOGAL survey (Omont et al. in preparation)
is a large set of Mid-IR images, which have been observed with
the ISOCAM camera on board the European satellite ISO, using filters centred
at 7~$\mu$m (LW2, LW5 and LW6) and 15~$\mu$m (LW3 and LW9),
with a pixel scale of usually 6\arcsec\ and sometimes
3\arcsec\ field of view. The spatial resolution of the ISO satellite was
$\sim$3\arcsec\ at 7~$\mu$m and $\sim$6\arcsec\ at 15~$\mu$m.
In total, more than 15 square degrees of sky have been mapped,
mostly in the Galactic disk, with galactic latitude in the
$\pm$1$^{\circ}$ range. A few fields at high galactic latitude were also
included in the survey. Since we don't expect many YSOs in these regions far
from the Galactic Plane, they will be used in the present analysis mainly as
checks of the results on identification of YSOs in the Galactic Plane.

A point spread function (PSF) fitting algorithm has been used to
extract the sources from the images (see the ISOGAL Explanatory Supplement,
Schuller et al. in preparation for details
on the used procedures and on the quality checks that have been achieved).
Crowding and blending effects can become significant and result in misleading
photometry for relatively faint sources in fields with high source density,
therefore strong photometric cuts have been applied to the initial catalogue
of extracted sources, primarily based on the results of artificial
stars simulations (see Schuller et al. in preparation). Then, a cross-identification
between 7~$\mu$m and 15~$\mu$m positions has been done using a correlation
radius of two pixels, to avoid missing identifications for slightly
extended sources, as can be the case for strong YSOs, for which the
7~$\mu$m and 15~$\mu$m ISOGAL positions may refer to different peaks.

The internal consistency of the photometry is guaranteed by the use of a fixed
PSF to extract all the sources with a given observational setup (filter and
pixel size), and an absolute flux density calibration has then been performed
by comparing the extracted fluxes with the predicted ones (M. Cohen, private
communication) for four stars in the Hipparcos catalogue with known spectral
types and luminosities, and by comparison between the ISOGAL and MSX (Price et
al.~\cite{MSX}) magnitudes for a large number of point sources.  As a result,
the photometry should not be biased by more than 0.1 magnitude, except maybe
0.2 magnitude for the faintest sources in the catalogue, with a typical
standard deviation of 0.15--0.2 magnitude,
and the completeness level should remain
above 50\% down to the used limit magnitudes (see Section~\ref{pscphotcuts} and
Schuller et al. in preparation for
details). This limit ranges from magnitude 8.2 near the Galactic Centre to 10.1
in low-density fields at 7~$\mu$m, and from 7.0 to 8.8 at 15~$\mu$m. The
corresponding limit flux densities, using the zero point fluxes as reported in
Paper~II, range from 6~mJy to 28~mJy at 15~$\mu$m, and from 8~mJy to 35~mJy at
7~$\mu$m, where the effects of crowding are more severe.

%

The total area covered by ISOGAL is divided into three types of fields:
1) fields of
type A observed only at 7~$\mu$m, 2) fields of type B observed only at
15~$\mu$m, 3) fields of type C observed at 7 and 15~$\mu$m; only in this case a
non-detection in either band can be used as an upper limit, with the exception
of extended sources as explained in Section~\ref{PSC_limit}.
The area covered only at 7 (2.1 deg$^2$) or only at 15~$\mu$m (2.7 deg$^2$)
is non-negligible with respect to that covered at both wavelengths (10.7 deg$^2$).
In the A and B fields the [7]--[15] colour criterion
cannot be used. Considering that only the reddest sources are
good  YSO candidates, we will retain only  the C fields where the red and bright
at 15~$\mu$m sources might be YSOs in our analysis.
The regions of the  Galactic Plane covered by ISOGAL are located in a
randomly sampled strip with latitude extension of about $\pm$~1$^{\circ}$
and  extending from l~=~-60$^{\circ}$ to l~=~140$^{\circ}$.
The distribution in galactic coordinates of the observed fields
of type C is shown in Fig.~\ref{fields}.

\subsection{The DENIS Data}
\label{Denis_data}
The sources in the $\delta\leq$+2$^{\circ}$ range have also been
associated with the DENIS database (Epchtein et al.~\cite{DENIS})
with a search radius roughly equal to one ISOCAM pixel. The DENIS source
density has been limited by cuts based on the K$_{s}$ magnitudes, in order
to limit the fraction of spurious associations to a few percents.
Most of the DENIS data come from special dedicated DENIS observations,
for which the source extraction has been performed by the ISOGAL team at
Observatoire de Paris (G. Simon, private communication),
and the astrometry has been matched with
the USNO-A2.0 catalogue, which has an accuracy of typically 0.25''
(Stone et al.~\cite{USNO}). As the internal astrometric accuracy of
the DENIS catalogue is better than 0.5'' (Epchtein et al.~\cite{DENIS2}),
the resulting uncertainty of the DENIS positions should be at
most 0.7''.
For the ISOGAL fields within the DENIS coverage, the mid-infrared astrometry
has been tied to the more accurate near infrared survey.
Nevertheless, the cross-identification
between ISOGAL and DENIS sources has shown that the offset between the
two databases can reach 10\arcsec, mainly due to the uncertainty in the
position of the filter wheel in ISOCAM, so that the astrometric uncertainty
in fields without DENIS counterpart (in galactic coordinates, this
corresponds roughly to l$\geq$+35$^{\circ}$) can reach this value.
The associations with DENIS data may also be limited for the
brightest sources, because of saturation problems, which occurs at
K$\approx$6. Such strong sources could not be properly extracted
and included in the DENIS catalogue.

\subsection{Intrinsic limitation of the ISOGAL-PSC}
\label{PSC_limit}
One basic limitation to the list of YSO candidates that we shall present comes
from the ISOCAM observations themselves:
in order to avoid saturation effects, the ISOGAL fields were
designed to exclude strong IRAS sources. No IRAS
source with F$_{12\mu m}$ $\geq$ 6~Jy  should be present in the ISOGAL fields
observed with broad band filters LW2 and LW3 (however, such a condition
has been relaxed
in a few regions of star formation and in the vicinity of the Galactic Centre,
observed with the narrower band filters: LW5, LW6 and LW9).
Consequently, the input lists will contain only
sources with relatively low flux densities,  not accessible by IRAS, namely
from the $\sim$ 10~mJy limit of the ISOGAL data at 15~$\mu$m  to the upper
limit quoted above, and only a few brighter sources (up to 35~Jy at 15~$\mu$m)
in the ISOGAL fields which were observed with narrow filters.

Another intrinsic limitation of the PSC is that by definition it contains
only point sources and sources with small extension,
while YSOs might also be significantly
extended, as it will be discussed in Section~\ref{sigbwhz}
and~\ref{suchiimorph} from
the inspection of ISOGAL images of radio identified YSOs.
This may introduce an error on the 7 and 15~$\mu$m
flux densities of extended YSOs, and aperture photometry on slightly
extended sources has shown that the PSF extracted magnitude can be
underestimated by about one magnitude. Moreover, sources that cannot be 
properly modelled with an unresolved component (i.e. extended sources)
are rejected by
the source extraction algorithm and don't appear in the PSC.

\begin{figure*}[tp]
\centerline{\psfig{figure=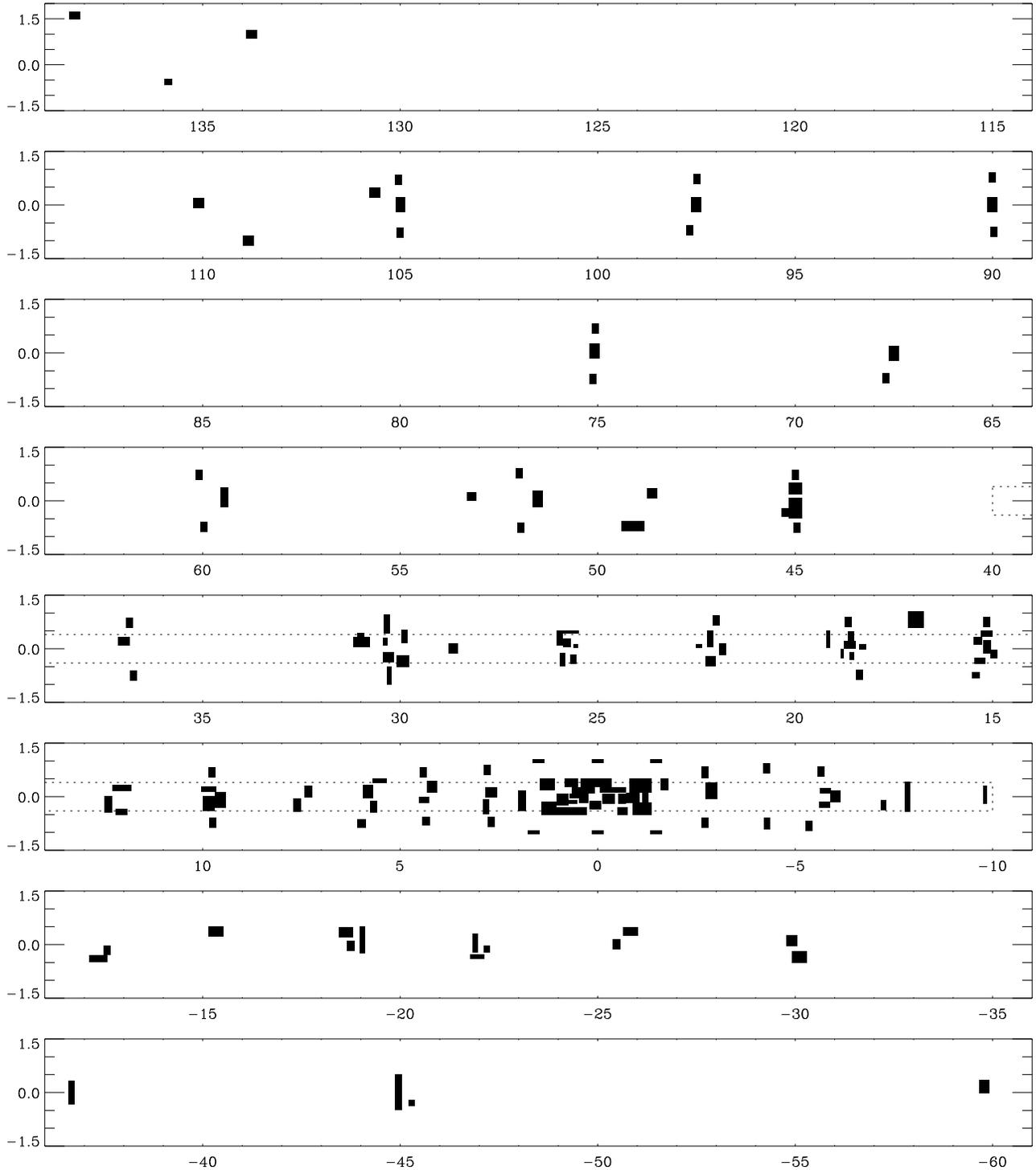,width=17cm}}
\caption[]{Galactic map of the ISOGAL FC fields. The graduations along the axis
are galactic coordinates, in degrees. The overlaid dotted line
shows the formal limits of the BWHZ survey.}
\label{fields}
\end{figure*}

\subsection{Differences in the ISOGAL data used in Paper~II}
\label{sigpscdiff}

In this section we want to stress the main differences in the ISOGAL PSC
with respect to the preliminary version that we used in Paper~II.
Two main improvements have been achieved, concerning the photometric calibration
and the reliability of the sources, which is related to the applied
photometric and geometric cuts.

\subsubsection{Photometric calibration}
\label{ig_phot_calib}

A careful analysis of the biases in the extracted fluxes, mainly due to a non
complete stabilisation of the signal because of short integration times, and to
an incomplete recovery of the flux density inherent to PSF fitting procedures,
as well as an absolute calibration with standard stars and with other MIR
surveys resulted in a correction of all extracted magnitudes by a constant
-0.45 magnitude offset (see Schuller et al. in preparation). This means that the colour
criterion that we derived in Paper~II is still relevant, but the magnitudes
from the present ISOGAL PSC are 0.45~mag brighter than in the preliminary
version. As we also decided to limit our analysis to bright YSO candidates, in
order to reduce the contamination of our sample by evolved stars, we will
derive new criteria based on the [7]--[15] colour and on the 15~$\mu$m magnitude,
the latter one being not directly comparable with the results of Paper~II.

\subsubsection{Geometric cuts}

The ISO observed rasters have saw-tooth shaped edges due to the different
orientations of the raster (along the galactic axis) and of the individual
ISOCAM images (aligned with the satellite axis, which match the equatorial
referential). Additionally, the 7 and 15~$\mu$m observations of a given field
do not perfectly match. Moreover, the source extraction procedure cannot work
properly for a source too close to the edge of the image, because it needs a
frame of pixels at least as large as the PSF representation. Therefore,
conservative limits have been derived for each field, given the constraints
that the retained area must be completely observed and at least at 2 pixels
from the edges at both wavelengths, and should be rectangular regions aligned
along the galactic coordinates.  The public version of the ISOGAL PSC contains
all the sources inside and outside these limits, but as the colour criterion
cannot be trustfully applied in the outside region, we will limit our analysis
to the formal field limits.

\subsubsection{Photometric cuts}
\label{pscphotcuts}
The other drastic difference in the ISOGAL PSC comes from the results of
artificial source simulations (details can be read in Schuller et al. in preparation)
whose purpose was to limit the published catalogue to reliable sources (this
means, to discard spurious sources and sources with too much biased photometry)
and to provide a catalogue about 50\% complete down to its faint end. The net
result is that about 25\% of the extracted sources have been discarded from the
catalogues before the cross identification between 7 and 15~$\mu$m took place,
this fraction being highly dependent on the source density in the field and
ranging between zero and almost 50\%.

\section{The BWHZ sample and its FIR properties}
\label{obs}

\subsection{The 6 cm BWHZ Galactic Plane radio survey}
\label{HII}

BWHZ observed the inner Galaxy ($\vert$b$\vert <$~0.4$^{\circ}$,
l~=~350$^{\circ}$-40$^{\circ}$) at 5~GHz with the VLA  down to a limiting
sensitivity of 2.5~mJy and with an angular resolution of $\sim$4\arcsec.
Extended structures (greater than 100\arcsec) were filtered out by the
instrument and reduction procedure.
The same region had also been observed at 1.4 GHz by
Zoonematkermani et al.~\cite{ZHB90}, Becker et al.~\cite{BWMHZ90},
White, Becker \& Helfand~\cite{WBH91} and  Helfand et al.~\cite{HZBW92}
with a maximum angular resolution of 5\arcsec.

The 5 GHz survey resulted in a catalogue of 1272 small diameter
discrete sources of which
$\sim$450 were tentatively classified as UC HII regions
using different classification criteria.

\begin{figure}[tp]
\centerline{\psfig{figure=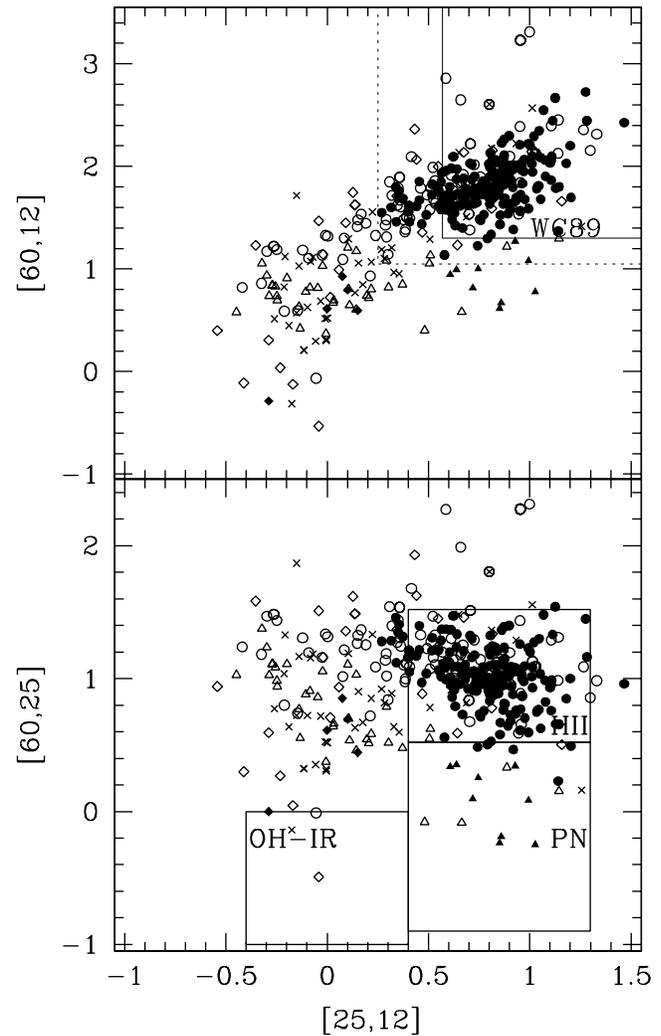,width=8.8cm}}
\caption[]{
[60,12] vs. [25,12] (top) and [60,25] vs. [25,12] (bottom) FIR
colour-colour diagrams of all the IRAS sources associated with radio
sources in BWHZ. Circles represent sources classified as UC~HII or candidate
UC~HII, triangles PNe or candidate PNe, diamonds are unclassified or
ambiguous sources, crosses are low-probability association sources (see BWHZ
for association and classification criteria); filled symbols are sources with
good fluxes in all three IRAS bands, open symbols have upper limits in at least
one band. In both diagrams:
[$\lambda_i$,$\lambda_j$]$=$Log$_{10}$(F$_{\lambda_i}$/F$_{\lambda_j}$).
In the top diagram,
the full line  box defines the boundaries of UC~HII
regions proposed by WC89, the dotted box
defines the relaxed colour criteria used to classify candidate UC~HII (see text).
In the bottom diagram the three labelled boxes define the regions of the
colour-colour plane occupied by OH-IR stars, Planetary Nebulae and HII regions
(Eder et al.~\cite{Eea88}; Pottasch et al.~\cite{Pea88})}
\label{colcol}
\end{figure}

The largest and most reliable number of identifications  are 291 radio sources
matching 246 IRAS sources with  UC HII regions  colours
(Wood \& Churchwell~\cite{WC89}, hereafter WC89) plus  17 radio-IR
matches with similar colours,  slightly
outside the formal survey boundaries, for a total of 308 small diameter
radio sources (and 263 IRAS sources). The strict WC89
colour-colour criteria were used for
``secure'' identification (Log$_{10}$(F$_{60{\mu}m}$/F$_{12{\mu}m})
\geq$ 1.3 and Log$_{10}$(F$_{25{\mu}m}$/F$_{12{\mu}m}) \geq$ 0.57)
and slightly larger criteria for ``candidate'' HII regions
(Log$_{10}$(F$_{60{\mu}m}$/F$_{12{\mu}m}) \geq$ 1.05 and
Log$_{10}$(F$_{25{\mu}m}$/F$_{12{\mu}m}) \geq$ 0.25).
Also sources that satisfy
the WC89 colour-colour criteria as well as  that
for planetary nebulae  but have   F$_{60{\mu}m} \geq$ 60 Jy
(or  F$_{60{\mu}m} \geq$ 100 Jy
using the ``candidate'' criteria) were classified as UC HII regions.
The situation is  summarised in  Fig.~\ref{colcol} where the [25--12]-[60--12]
and [25--12]-[60--25]
colours of the 263 IRAS sources with one or more radio components are reported.


Most of the 263 IRAS sources (74\%) are in the WC89 box, a consistent
fraction (18\%) in the ``candidate'' region and a minor fraction (8\%)
outside. In this last case  all sources have at least an upper limit in one
IRAS band. However, we note that only 159 IRAS sources of the  BWHZ sample
fully satisfy the WC89 criteria, i.e. no upper limits at 25 and 60~$\mu$m and
colours within the WC89 box.

The fact that more than one {\it compact}
radio component is associated with a single IRAS
source is not a new finding. HII regions are usually
very structured and often  have
several  small diameter components all clustered in an area
of few tens of arcsec, i.e. well within the resolution of
the IRAS beam. In some cases they are indeed independent HII regions and
in VLA radio surveys of  HII regions (Wood \& Churchwell~\cite{WC89s} and
Kurtz, Churchwell \& Wood~\cite{KCW})
the percentage of irregular or
multiple-peaked sources is $\sim$ 18\%, in very good agreement with the
present situation: (308 -263)/263 = 17\%.

Two other criteria are used by BWHZ
to reach the $\sim$450 putable HII regions:
the radio spectral index and the galactic latitude distribution.
There are
84 sources which were observed both at 5 and 1.4 GHz
and which have a thermal spectral index but do not have an IRAS counterpart
(out of 220 detected at both wavelengths and with a thermal spectral
index).
Their galactic latitude distribution  peaks toward the Galactic Plane and
further supports their Galactic nature.

Finally, the galactic latitude distribution  of all the
remaining unidentified radio sources (i.e. without an IRAS counterpart or
with no  spectral index
information)  still shows a clear excess toward
the Galactic Plane and cannot be background extragalactic sources.
This leads to an estimated  $\sim$90 additional thermal
Galactic sources, which however cannot be identified individually with
only the BWHZ observations.

\subsection{Relation F$_{radio}$-F$_{FIR}$}
\label{srfirflr}

In the present section we examine the expected relation between radio
and MIR flux densities for UC HII regions, both from observations and theory.
This relationship will be needed in the following as an additional tool
to discriminate between late type stars and YSO candidates in the ISOGAL PSC.
For this purpose we shall use the 263 IRAS sources in the BWHZ list of
UC HII regions. When more than one radio component is associated with the
same IRAS source, we simply added up the radio flux densities.

\begin{figure*}
\resizebox{15cm}{!}{ \rotatebox{0}{\includegraphics{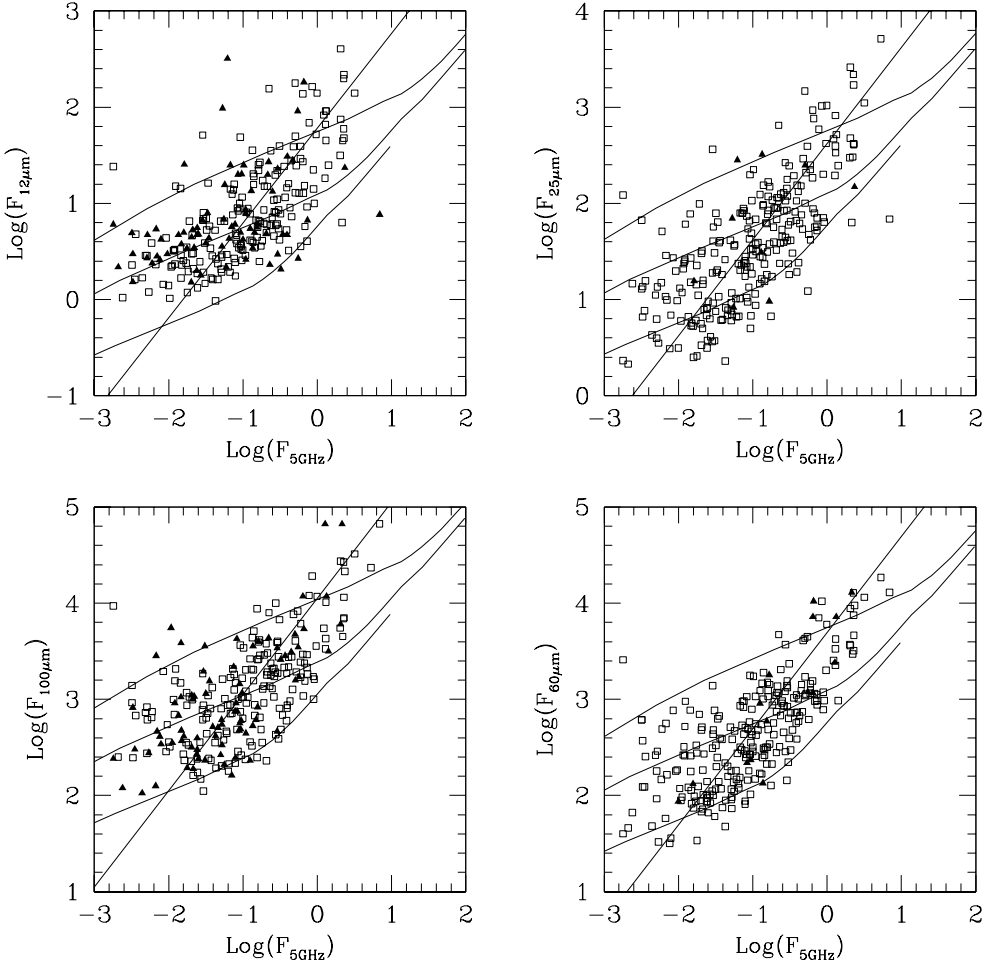} }}
\caption[]{
F$_{5GHz}$ versus F$_{FIR}$ (both in Jy) for the BWHZ sources associated with an IRAS
selected UC~HII. The expected relation between F$_{5GHz}$ and F$_{FIR}$
for Log(D$^2$/kpc$^2$) = 0 (upper), 1 (middle)
and 2 (lower) are overlaid, the linear relations of Table~\ref{mean} are 
also shown.
Open squares represent good IRAS detections, filled
triangles sources with an upper limit.
}
\label{teoff}
\end{figure*}

Fig.~\ref{teoff} shows F$_{5GHz}$ versus F$_{FIR}$
for the  263 IRAS selected UC HII regions of the BWHZ sample.
The spread of the points is very large and comes from the  fact that we
are mixing  stars of different spectral types, but a general correlation
between  the two fluxes can be noted.
In two plots,  the 12~$\mu$m
and 100~$\mu$m, a clear deviation from a linear correlation can also be
noted.
In Table~\ref{mean} we give the mean value at each
IRAS band of the FIR over the radio emission (both in Jy), which shows a
very rapid increase with wavelength up to a maximum at 100~$\mu$m
(more than four orders of magnitude).
The linear interpolation at 15~$\mu$m indicates that F$_{15\mu m}$
is about two orders of magnitude greater than F$_{5GHz}$.
A similar value was derived by Churchwell et al.~\cite{CWW90}
from a model fit of the SED of G5.89-0.39, a dust cocoon surrounding an
O6 ZAMS star.
While this flux ratio is amply compensated by the lower sensitivity of IRAS
with respect to radio surveys, with ISO, which has a sensitivity comparable
to that of radio surveys, we should be able to observe many YSOs
which are undetected in the radio.

\begin{table}
\caption[]{\label{mean}Mean FIR to 5 GHz flux ratios}
\begin{tabular}{rcr}
\hline
$\lambda$/$\mu$m &  $\langle$ LogF$_{FIR}$ - LogF$_{5GHz}$ $\rangle$
 & $\sigma$\\
\hline
12 & 1.86 & 0.52 \\
25 & 2.62 & 0.53 \\
60 & 3.70 & 0.51 \\
100 & 4.05 & 0.57 \\
\hline
\end{tabular}
\vskip 0.3cm
\end{table}

The radio emission from a spherical, homogeneous, optically thin HII region
is proportional to the stellar ionising photon flux N$_L$.
At 5 GHz, assuming an electron temperature of 10$^4$ K, we have:

F$_{5GHz}$/Jy = 1.1285 10$^{-47}$ N$_L$ $\xi$ [D/kpc]$^{-2}$

where $\xi$ is the fraction of N$_L$ that goes into ionization of the gas
(in an ionization bounded dusty HII region (1~-~$\xi$)N$_L$ photons
are absorbed by dust), and D is the distance.
According to WC89,  $\xi$ is, on the average, very small, about 16\% for a
sub-sample of 19 sources for which $\xi$ was estimated.
Such a small value of $\xi$ is explained by
WC89 as an evolutionary effect, i.e. the
sample selects preferentially young HII regions ionised by stars
with very early spectral type. In fact, dust absorption in the Lyman
continuum increases with electron density (hence with youth of the region)
and is stronger for earlier
spectral type stars (see e.g Figure 2 of Felli~\cite{felli}).
However, a more simple alternative solution for such small values of $\xi$
could be the result of  missing radio  flux density, i.e. that of
extended envelopes surrounding the more compact structures,
which are filtered out in the interferometric observations.
This is probably the most plausible explanation, as also
pointed out by Kim and Koo~(\cite{kk}) from a comparison of radio flux densities
of UC HII regions obtained with high and low resolution. For our
purposes we shall use $\xi=1$.

The FIR emission is proportional to the total stellar luminosity L.
In order to derive the constant of proportionality between F$_{FIR}$ and L
we use the mean value of the ratio L/(F$_{FIR}$D$^2$)
for the sources of the WC89 sample that do not have upper limits in the
IRAS bands. The values of $\langle$ L/(F$_{FIR}$ D$^2$) $\rangle$
in the four IRAS bands are given in Table~\ref{const}.
At 10~$\mu$m  Walsh et al.~(\cite{wbbn}) from 5 UC HII regions find
$\langle$ L/(F$_{FIR}$ D$^2$) $\rangle$ = 2080 L$_{\odot}$ Jy$^{-1}$
for D = 1 kpc, with a rather large uncertainty.
A linear fit to all these values gives
$\langle$ L/(F$_{FIR}$ D$^2$) $\rangle\sim$300 L$_{\odot}$ Jy$^{-1}$
kpc$^{-2}$ at 15~$\mu$m.

The expected values of F$_{5GHz}$ and   F$_{FIR}$
for UC HII regions ionised by single stars of different spectral type
can be derived from the tabulation of Panagia~\cite{pan} which gives
N$_L$ and L for stars of early spectral type (from O4 to B3);
ZAMS values were used.
Ionization of the UC HII region by a stellar cluster rather than by
a single star have also been considered (Walsh
et al.~\cite{wbbn}) and the relationships between F$_{5GHz}$
and F$_{FIR}$ are slightly different.

\begin{table}
\caption[]{\label{const} Mean value of L/F$_{FIR}$D$^2$}
\begin{tabular}{rcr}
\hline
$\lambda$/$\mu$m &  $\langle$ L/(F$_{FIR}$ D$^2$) $\rangle$/L$_{\odot}$ Jy$^{-1}$
kpc$^{-2}$
 & $\sigma$\\
\hline
12 & 332.2 & 444.0 \\
25 & 32.47  & 22.86 \\
60 & 3.339 & 0.439 \\
100 & 1.691 & 0.368 \\
\hline
\end{tabular}
\vskip 0.3cm
\end{table}

In Fig.~\ref{teoff} we show for the four IRAS bands the expected
relation between  the 5 GHz flux density and the FIR flux for single stars,
overlaid on the observed points for the BWHZ sample.  The match with the
observations is satisfactory. The three curves are for
Log(D$^2$/kpc$^2$) = 0, 1 and 2 from top to bottom.
The lower curve (10 kpc) delimits well the lower distribution of points,
an effect
due to a combination of the edge of our galaxy  and the sensitivity limits
of radio observations. In the single-star approximation, 
points above the upper curve should be closer than 1 kpc.

It is clear from this figure that the observed spread of points and the shape
of the correlation are  due
to a combination of different distances, which moves the points
along the bottom-left top-right diagonal, {\it and}  of the presence
of stars of different spectral type and luminosity,  which
moves the points along the drawn lines.  These lines become steeper and closer
to the diagonal,
and hence the two effect are indistinguishable for
very early spectral types (earlier than O7).

\section{ISOGAL counterparts to BWHZ sources}
\label{scrco}

In this section we will use the comparison of the ISOGAL-PSC data
with the radio continuum 5~GHz survey of BWHZ
to check the criteria developed in papers~I and II
to separate in the ISOGAL-PSC the YSOs from the much larger
population of Post-MS stars. The comparison with the radio survey
will allow to refine the criteria to select relatively luminous
YSOs, those capable to ionise the surrounding material, and to produce
a detectable radio continuum flux.

The design of the Galactic Plane coverage in the BWHZ and ISOGAL surveys
were bound to very different instrumental and scientific constraints.
The radio survey was aimed at the complete, and uniform coverage of a
large longitude and narrow latitude range of the portion of the inner
Galactic Plane visible from the Very Large Array, with the aim of
detecting a complete sample of young massive stars, through the radio
continuum emission of the surrounding ionised gas. As a byproduct, a
significant number of planetary nebulae, supernova remnant and a minor
fraction of radio stars and background radio sources were detected.
The ISOGAL survey was mainly
designed to study the galactic distribution and properties of
late type, Post-MS stars, exploring a large longitude
range, but with non-continuous, non-uniform and uneven sampling of the
plane. Additionally, saturation limits of the ISO detectors forced
to systematically avoid bright MIR sources (see Sect.~\ref{PSC_limit}).
The different area coverage of the two surveys is exemplified in
Figure~\ref{fields}.
A comparison of the two samples should check that no bias is
introduced by the different sampling, except for the obvious note
that bright sources will be absent from the ISOGAL dataset.

\subsection{BWHZ sources within ISOGAL fields}
\label{sbwhzinig}

Of the 1272 sources in the BWHZ sample 138 fall within the boundaries
of the ISOGAL fields considered in this paper (see Sect.~\ref{sigcat}).
Most of these radio sources are unclassified, except for 3
supernova remnants (SNR) or candidate SNR, 34 sources with a high
probability of being associated with IRAS sources, and 4 additional
lower probability IRAS associations (BWHZ). Most (29) of the sources
associated with IRAS sources are classified as UC~HII or candidates UC~HII
by BWHZ, one is a planetary nebula (PN) and only 4 could not be classified.
Among the radio sources not associated with IRAS counterparts
(hence unclassified), the majority is expected to be UC~HII without
far infrared counterpart due to the incompleteness of the IRAS-PSC
in the Galactic Plane (BWHZ).
Following the discussion of BWHZ and the survey of Fomalont et
al.~(\cite{Fea91}) at the sensitivity limits of the radio survey we expect a
contamination of background extragalactic radio sources of $\sim$23\%,
and 1--2\% of unclassified non-UC~HII Galactic objects (radio stars, pulsars,
planetary nebulae, etc.). Thus, among the 138 radio sources within
ISOGAL fields, $\sim$35 of the 100 that are unclassified are not expected to be
YSOs (UC~HII regions).

\begin{figure}
\centerline{\psfig{figure=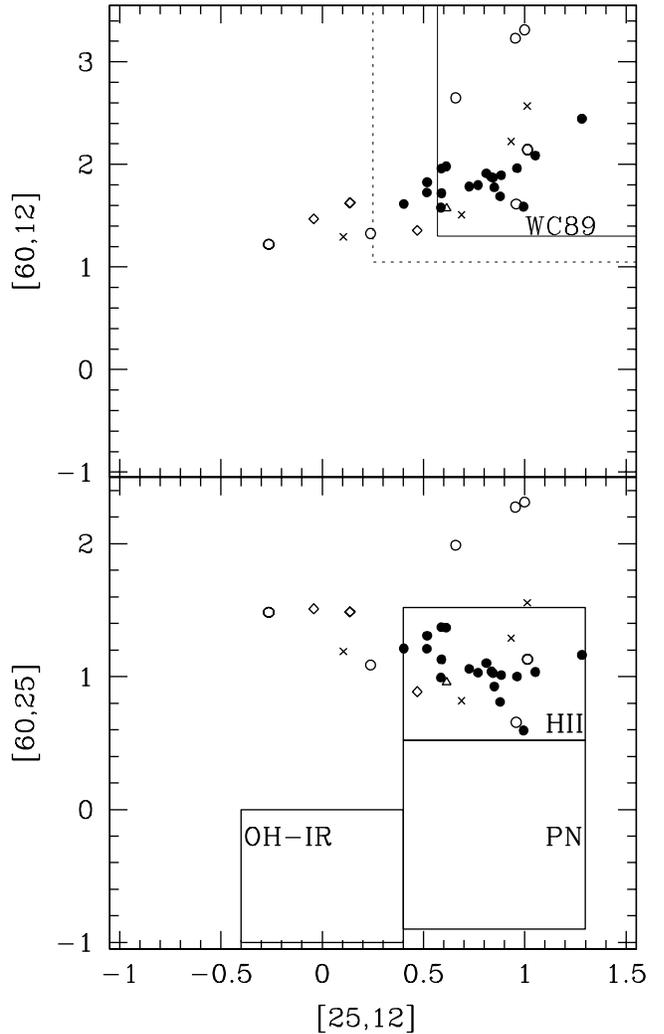,width=8.8cm}}
\caption[]{Same as Figure~\ref{colcol}, but for the sources within
the ISOGAL fields.}
\label{ffir_cc_inig}
\end{figure}

In Figure~\ref{ffir_cc_inig} we show the IRAS colour-colour plots
of the IRAS sources associated to BWHZ sources and within the
ISOGAL fields. Here we just point out that it appears an
unbiased random subsample of the total shown in Figure~\ref{colcol}.
The numbers of sources in the various classes are not significantly
different from what expected by scaling the original sample to
the area in common between the two surveys.

%
%

%

\subsection{Comparison of BWHZ sources with the ISOGAL PSC}
\label{sigbwhz}

In order to identify possible mid-infrared counterparts of the BWHZ sources, we
decided to use the criterion of positional coincidence.  Given the large number
densities of ISOGAL sources, especially in the inner regions of the Galaxy, the
result of this correlation and the reliability of the radio-ISOGAL source
association, or, alternatively, the number of spurious associations is expected
to be a strong function of the tolerance in the positional coincidence. The
approach we decided to follow is to define a tolerance, or searching radius,
based on the characteristics of the radio sources we are interested in, UC~HII
and candidates, which also constitute the vast majority of the radio sample.
Formal positional uncertainties in the radio and ISOGAL source lists are rather
small, generally less than 2\arcsec, however, source physical properties,
rather than positional uncertainties need to be used to set the association
criteria. UC~HII regions in the 5~GHz sample are generally extended (see BWHZ),
with typical sizes up to 15\arcsec. Moreover, the radio continuum and MIR
observations of UC~HII regions are not expected to trace exactly the same
material, the latter being sensitive to the YSOs circumstellar dust as opposed
to the ionised gas of the former. The comparison of near infrared and radio
continuum studies of compact HII regions has shown that at the resolution of a
few arcsec the radio continuum and infrared peaks may be separated by several
arcseconds (Testi et al.~\cite{Tea95}; \cite{TFPR98}; Watson et
al.~\cite{Wea97}; Feldt et al.~\cite{Fea98}; Walsh et
al.~\cite{Wea99};~\cite{wbbn}).  For these reasons we decided to
use a rather large association distance (15\arcsec) and to check the
reliability of the associations by comparing the properties of the real
BWHZ/ISOGAL-PSC matches with the results of extensive random simulations.

The details of the random samples realizations and analysis are discussed in
Appendix~\ref{app_rnd}. The main results of this analysis are that,
with the association radius we adopted, we could expect a fraction of
false associations that can be up to $\sim 50\%$. This probability,
however is a strong function of not only the association distance,
but also of the mid-infrared brightness and colour of the associated
sources. In practice the vast majority of the ``real'' radio sources
are found to be associated with sources that have mid-infrared
properties clearly different from those of ISOGAL sources associated to the
random samples. More quantitatively, only a relatively
small fraction ($\sim 19\%$)
of the ISOGAL sources that we have
associated with BWHZ sources do not meet at least
one of the high reliability criteria discussed in Appendix~\ref{app_rnd}.

\begin{figure*}
\centerline{\psfig{figure=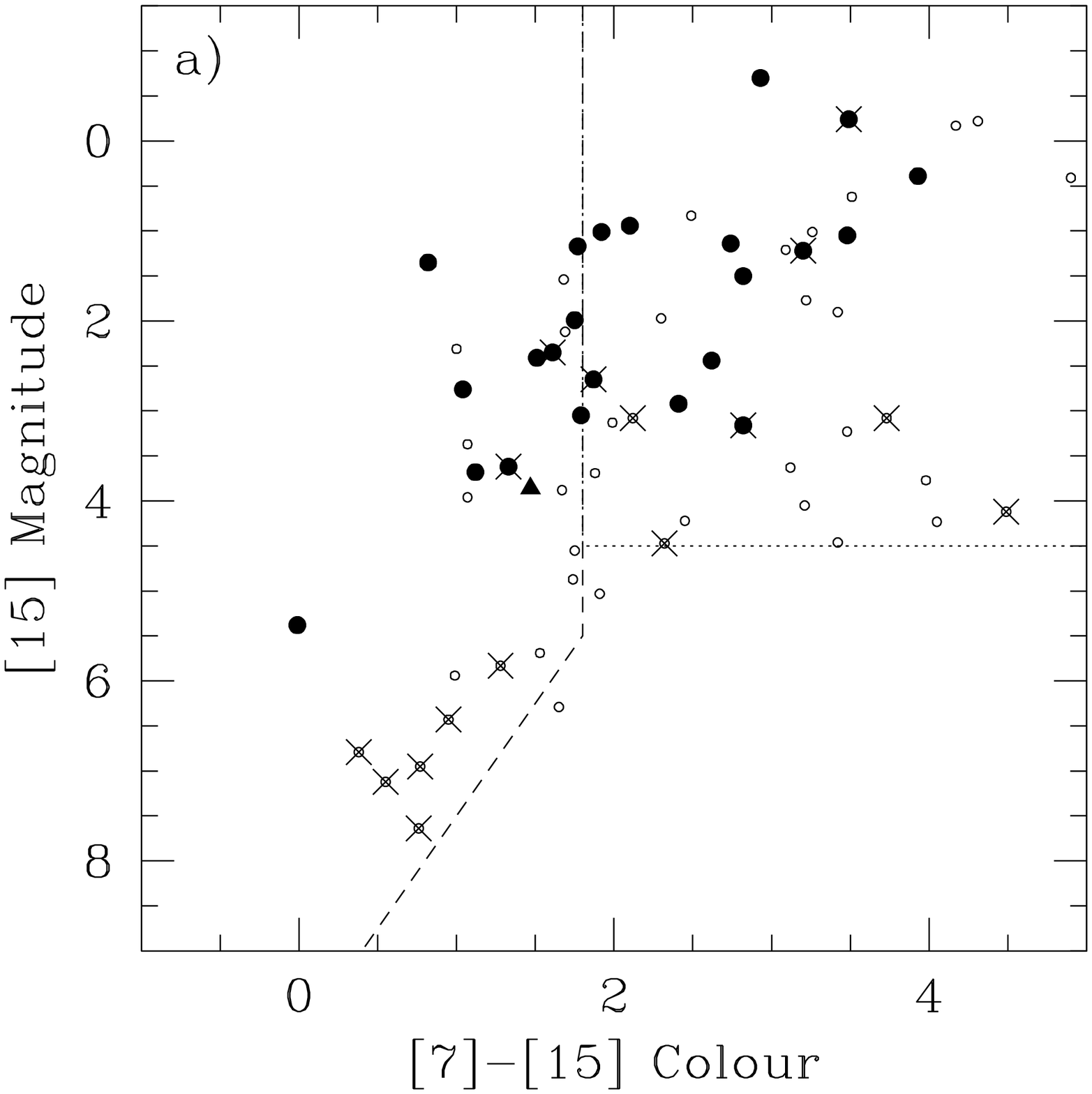,width=6cm}
            \psfig{figure=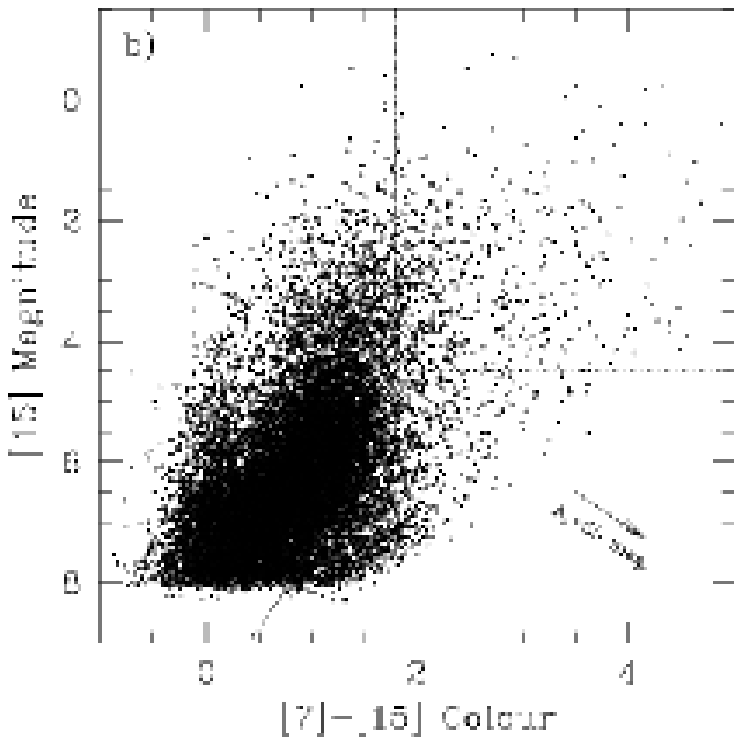,width=6cm}
            \psfig{figure=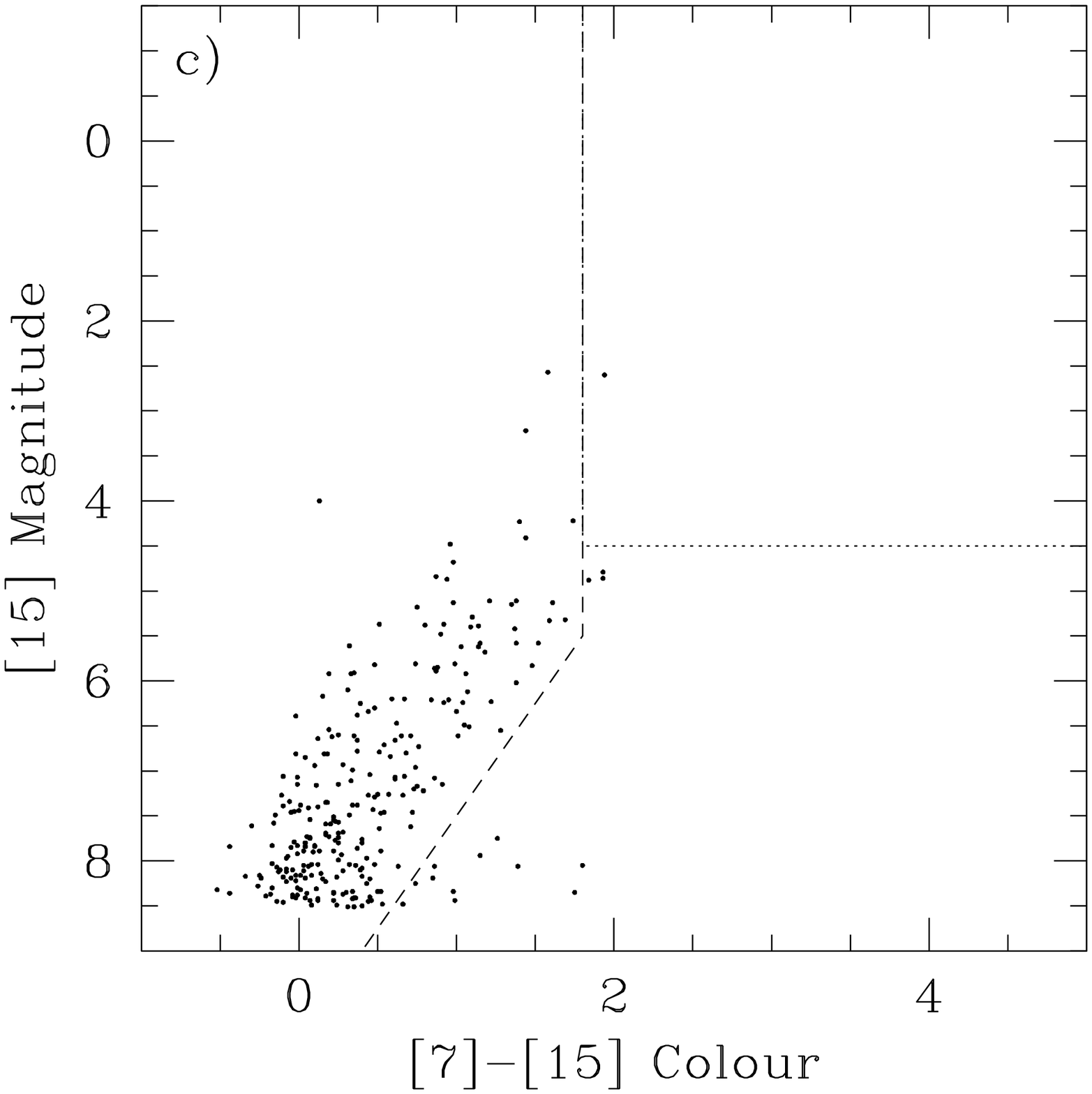,width=6cm}}
\caption[]{a) ISOGAL ([7]$-$[15],[15]) colour-magnitude diagram for the
sources associated with BWHZ radio sources and with good detections at
both 7~$\mu$m and 15~$\mu$m. Small open circles refer to unclassified radio
sources, either non associated with IRAS sources, with low association
probability or with ambiguous IRAS colours/fluxes (see BWHZ); filled circles
represent UC~HII or candidate UC~HII; the triangle marks the only PN in
the sample. Sources with association distance greater than 8 arcsec
are marked with a cross. b) ([7]$-$[15],[15]) colour-magnitude diagram for
all the ISOGAL sources within the BWHZ survey region. c) ([7]$-$[15],[15])
colour-magnitude diagram for a compilation of ISOGAL fields at high galactic
latitude: all fields with $|b|\ge 3^\circ$, except those in the Magellanic
Clouds. In all diagrams the dashed line shows the criteria used to derive
YSO candidates in Paper~II, the dotted horizontal
line shows the more conservative
limit on [15] adopted in this paper (see section \ref{sselc}).
The extinction vector for A$_{\rm V}$=50~mag is shown in panel b
(assuming the extinction law provided by Jiang et al., private communication).
}
\label{fcmd_ig_bwhz}
\end{figure*}

Of the 138 BWHZ sources within ISOGAL fields 96 can be associated with
ISOGAL-PSC sources within the adopted association distance.
Of the classified radio sources 26 out of the 29 UC~HII regions (or candidate)
and the only PN in the sample are associated with an ISOGAL source.
None of the radio sources classified as SNR is found to be associated with
an ISOGAL point source.
In Figure~\ref{fcmd_ig_bwhz}a we show the ISOGAL ([7]$-$[15],[15]) colour
magnitude diagram for the ISOGAL sources with good detection in both
filters (65 of 96 sources). Filled circles represent
UC~HII or candidate UC~HII; the triangle marks the only PN in
the sample; small open circles are unclassified radio sources, either
non associated with IRAS sources, low probability associations or
with ambiguous FIR fluxes and colours (see BWHZ). Sources with association 
distance greater than 8\arcsec\ are marked with a cross. We note that the
filled source in the lower-left of Figure~\ref{fcmd_ig_bwhz}a has also 
IRAS colours well outside the WC89 box, its classification as candidate
UC~HII (BWHZ) is rather dubious.

In Figures~\ref{fcmd_ig_bwhz}b and ~\ref{fcmd_ig_bwhz}c
we also show the colour-magnitude diagram of
all the ISOGAL sources within the BWHZ survey region and of all ISOGAL
sources with $|b|\ge 3^\circ$ (excluding those in the Magellanic Cloud).
The high latitude fields were chosen because we do not expect that
they contain a significant number of young stellar objects (if any at all).
In the figure we also show the lines defining the criteria devised in Paper~II
to select YSO candidates (dashed lines) as well as the revised criteria that we
will adopt in this paper (see Sect.~\ref{sselc}).

\begin{figure*}
\centerline{\psfig{figure=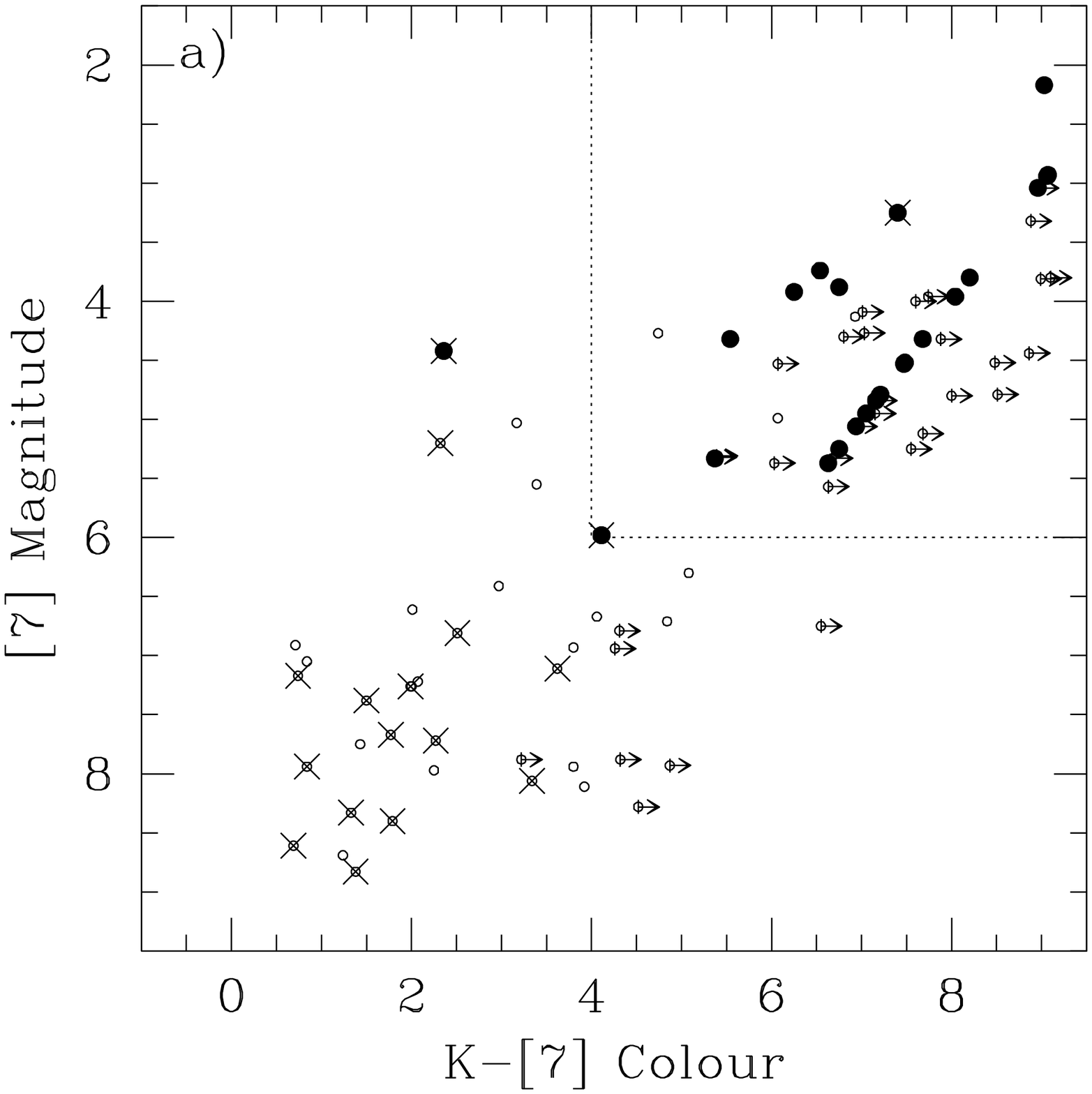,width=6cm}
            \psfig{figure=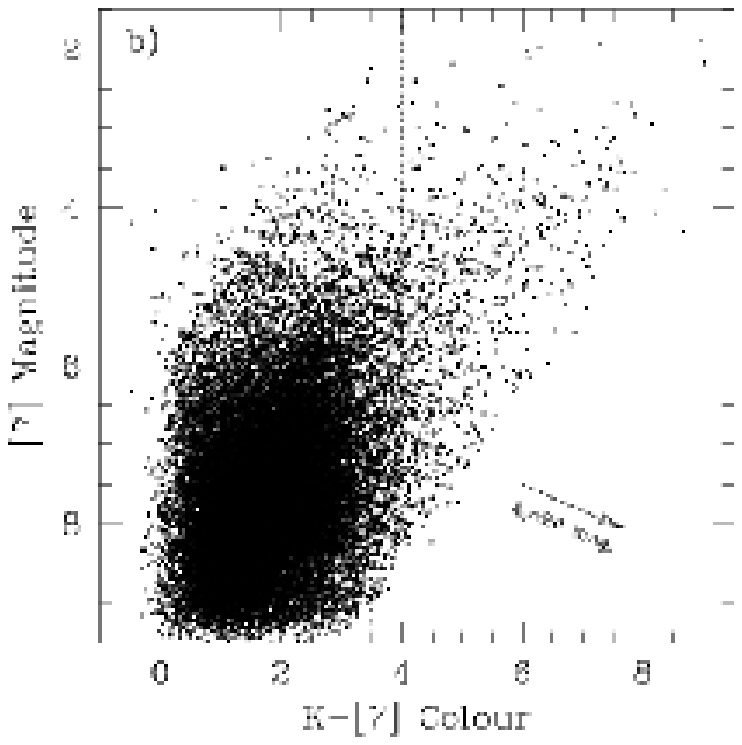,width=6cm}
            \psfig{figure=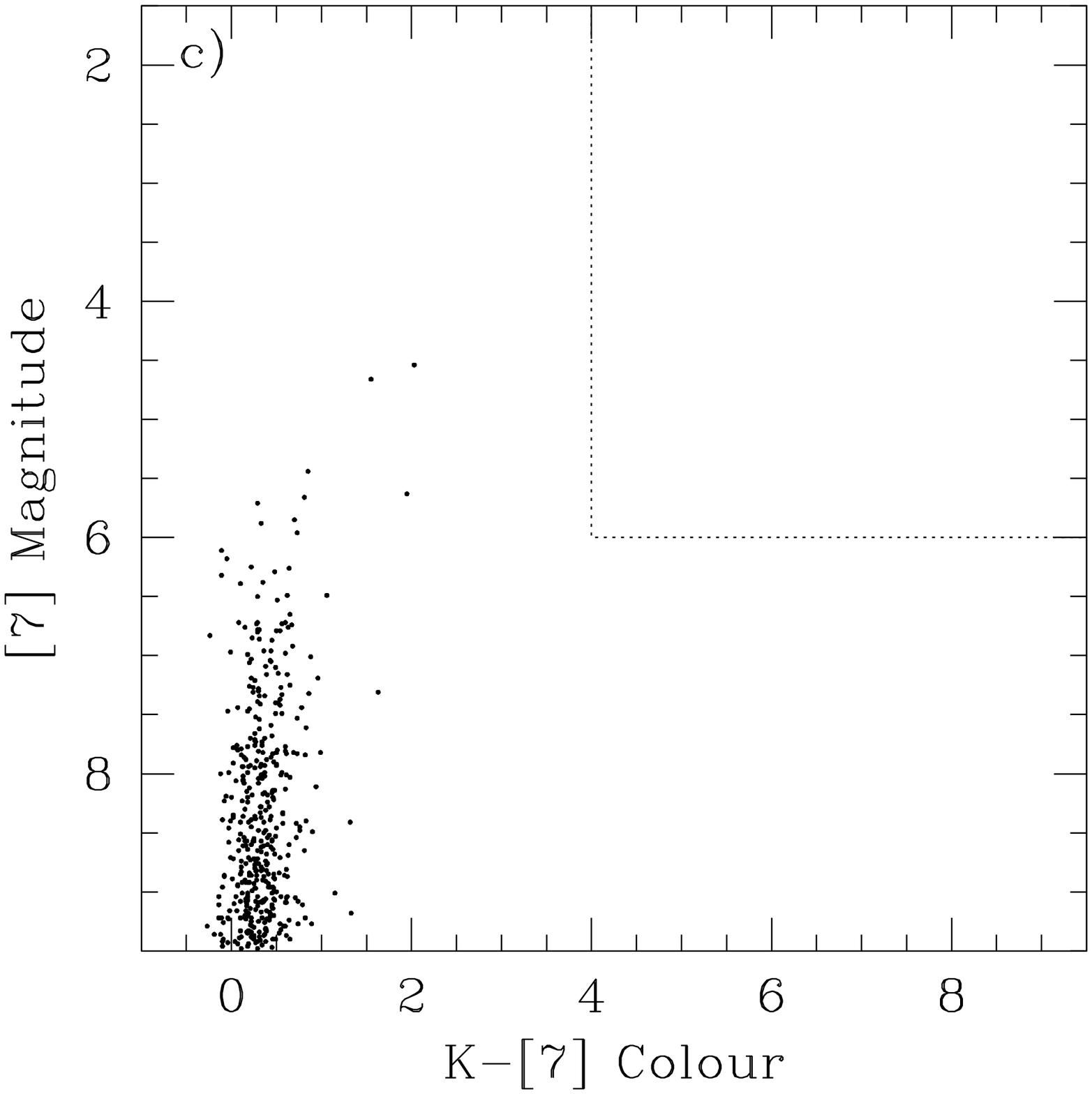,width=6cm}}
\caption[]{a) DENIS-ISOGAL (K$-$[7],[7]) colour-magnitude diagram for the
sources associated with BWHZ radio sources and with good detections at both
K-band and 7~$\mu$m; sources within fields observed by DENIS with good [7]
detection but no K-band detection are marked with a right arrow, their position
has been computed using the DENIS limiting magnitude for association
with ISOGAL (as defined in Omont et al. in preparation and
Schuller et al. in preparation). Symbols have the same
meaning as in Figure~\ref{fcmd_ig_bwhz}a, the triangle is missing because the
radio source classified as PN does not have a DENIS counterpart.  b)
colour-magnitude diagram for all the ISOGAL sources within the BWHZ survey
region. c) colour-magnitude diagram for a compilation of ISOGAL fields at high
galactic latitude: all fields with $|b|\ge 3^\circ$, except those on the
Magellanic Clouds.  The extinction vector for A$_{\rm V}$=20~mag is shown in
panel b (assuming extinction laws from Rieke \& Lebofski~\cite{RL85} and Jiang
et al., private communication).
}
\label{fdeniscmd}
\end{figure*}

With respect to Paper~II,
an important additional constraint that is available with the public version
of the ISOGAL catalogue is the near infrared photometry from the DENIS
survey (Epchtein et al.~\cite{DENIS}), for all ISOGAL fields in the southern
hemisphere (see Sect.~\ref{Denis_data}).
In Figure~\ref{fdeniscmd}a we show the (K--[7],[7]) colour-magnitude
diagram for all the ISOGAL sources with good detections at K and [7], and
associated with BWHZ sources. The symbols have the same meaning as in
Figure~\ref{fcmd_ig_bwhz}a, the PN does not have a DENIS counterpart,
even if it is within the fields with DENIS data, so the triangle is missing
from the figure. Sources with upper limit at K (including the PN) are shown 
with arrows. In Figures~\ref{fdeniscmd}b and~\ref{fdeniscmd}c we show
similar diagrams for all the ISOGAL sources within the BWHZ survey area
and all the high latitude ISOGAL fields (excluding the Magellanic Cloud).

As shown by the comparison between Figure~\ref{fcmd_ig_bwhz}a and
Figures~\ref{fcmd_ig_bwhz}b, \ref{fcmd_ig_bwhz}c, the radio selected UC~HII
fall preferentially in the top-right part of the ([7]--[15],[15]) plane and
form a subsample of ISOGAL sources with distinctly different colour-magnitude
properties with respect to the much larger population of Post-MS stars, which
are overwhelmingly concentrated in the lower-left region of the plane. As a
countercheck, in the high latitude fields (Figure~\ref{fcmd_ig_bwhz}c), where
no UC~HII are expected, almost all ISOGAL sources are located in the lower-left
part of the plot. Sources marked with crosses in Fig.\ref{fcmd_ig_bwhz}a,
which, as discussed in the Appendix, have a lower probability of being true
radio-ISOGAL associations, populate uniformly the colour-magnitude diagram,
supporting the hypothesis that a good fraction of these sources are not UC~HII.
Similar conclusions can be drawn from a comparison between
Figure~\ref{fdeniscmd}a and Figures~\ref{fdeniscmd}b and \ref{fdeniscmd}c. Also
in the (K--[7],[7]) colour-magnitude plot, the well identified UC~HII populate a
distinct part of the plot with respect to most of the ISOGAL sources.

\begin{figure}
\centerline{\psfig{figure=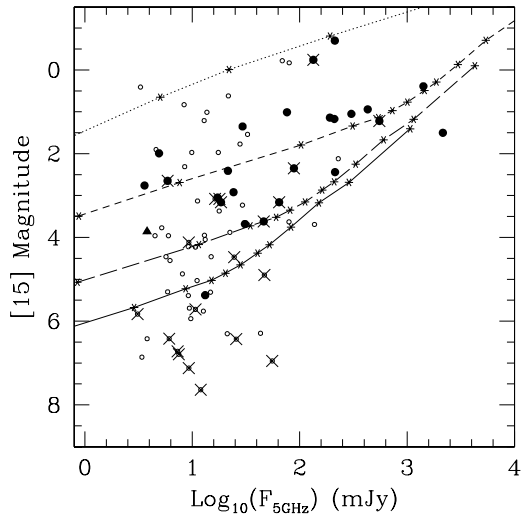,width=7.0cm}}
\caption[]{Radio continuum flux density versus [15] magnitude for the
BWHZ sources associated with ISOGAL sources with good 15~$\mu$m detections.
The symbols are as in Figure~\ref{fcmd_ig_bwhz}. The stars connected with lines
show the predictions for optically thin, spherical, and homogeneous HII regions
following the prescriptions of Panagia~(\cite{pan}) and
the ratios between F$_{15\mu m}$ and L as derived in Sect.~\ref{srfirflr}.
The lines correspond to
different distances from the Sun: 1 (dotted), 5 (short dashed),
15 (long dashed), and 30~kpc (solid).}
\label{ffluxm15}
\end{figure}

As an additional check on the reliability of the UC~HII region nature of the
ISOGAL sources associated with BWHZ sources, we can check whether the
[15] magnitudes are consistent with those predicted for UC~HII of a
given radio flux. In Figure~\ref{ffluxm15} we show
the radio flux-[15] relation for the BWHZ sources associated
with ISOGAL sources with good 15~$\mu$m detections. In the same figure we also
show the expected relation for optically thin, spherical and homogeneous
UC~HII, derived in Sect.~\ref{srfirflr}. The theoretical expectation is
shown for several distances ranging from 1 (dotted) to 30~kpc (solid).
The vast majority of BWHZ radio sources classified as UC~HII or candidate UC~HII
(black circles in Figure~\ref{ffluxm15}) are in good agreement with model
predictions, as opposite to the unclassified radio sources associated with faint
ISOGAL sources which fall well below the 15~$\mu$m brightness expected for UC~HII.
These sources would be consistent with the models assuming large extinction
values (exceeding 150 mags in the visual), which are not unreasonable for
star forming regions, but would imply [7]--[15] colours much larger than observed
(see Figure~\ref{fcmd_ig_bwhz}b). We believe that in most of these cases
either the radio source is not an UC~HII or the ISOGAL source associated
with the radio source is not physically related with it (i.e. is a spurious
association).
In sect.~\ref{sbwhzinig} we estimated that $\sim$35 of the BWHZ radio sources
are not YSOs, thus
it is not surprising that a fraction of ISOGAL sources associated
with BWHZ sources show infrared properties inconsistent with those expected
for YSOs. These sources should be disregarded when defining the selection
criteria to identify YSO candidates in the ISOGAL database.

\subsection{MIR morphology of UC~HII}
\label{suchiimorph}

\begin{figure*}
\centerline{\psfig{figure=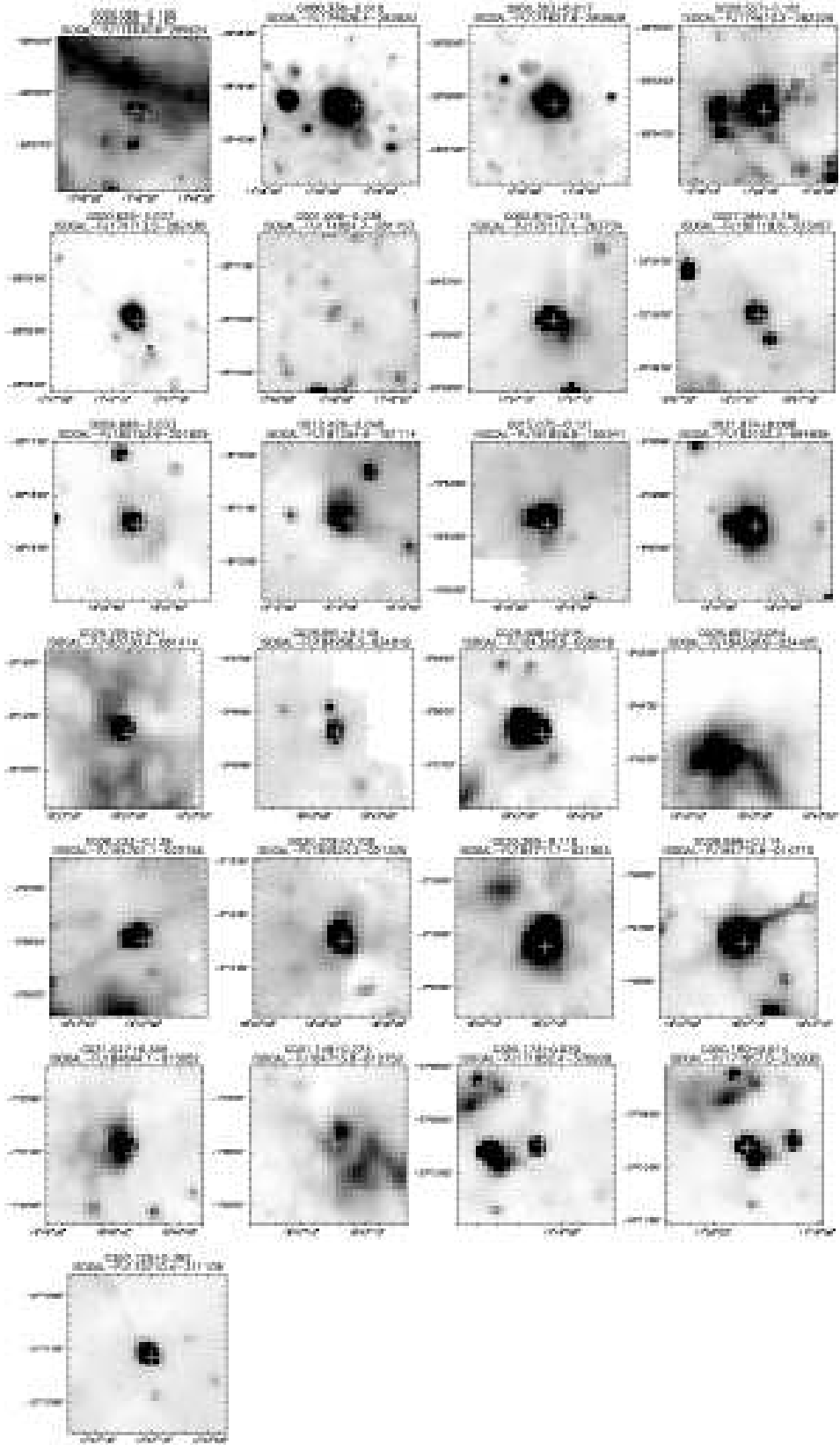,height=22cm}}
\caption[]{ISOGAL 15~$\mu$m maps of the fields surrounding the 26 BWHZ UC~HII
regions associated with an ISOGAL point source. The top left field
contains two radio sources, which are associated to the same ISOGAL source.
The crosses
mark the positions of the radio peaks. Each chart is 3$^\prime\times 3^\prime$
and it is centred on the ISOGAL source associated with the radio source(s).
On top of each panel are reported the names of the radio source(s) and ISOGAL
source.}
\label{fuchiidet}
\end{figure*}

\begin{figure*}
\centerline{\psfig{figure=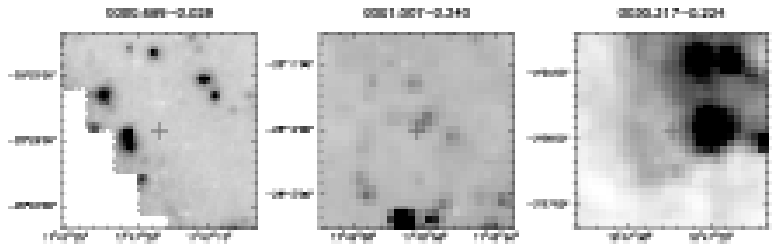,width=13cm}}
\caption[]{Same as Figure~\ref{fuchiidet}, but for the 3 UC~HII regions not
detected at either 7 or
15~$\mu$m. The charts are centred on the radio peak position.}
\label{fuchiiundet}
\end{figure*}

In Figure~\ref{fuchiidet} and~\ref{fuchiiundet} we show the
15~$\mu$m images of the UC~HII and candidate UC~HII (BWHZ) 
with (26 sources) and without (3 sources) an associated ISOGAL-PSC 
source. In each panel the cross marks the position of the 
radio peak (from BWHZ). Almost always the 15~$\mu$m image reveals a slightly 
extended nebulosity, in agreement with the resolved radio structure of the 
UC HII regions.


\begin{figure}
\centerline{\psfig{figure=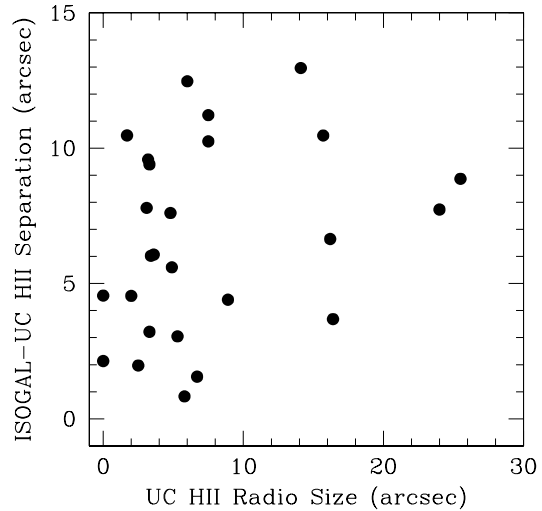,width=7.0cm}}
\caption[]{Distance between the radio and MIR peaks for the 26 UC~HII regions
associated with an ISOGAL-PSC source. }
\label{fuchiid}
\end{figure}

As discussed earlier, the slightly extended MIR and radio 
structure of the YSOs amply justifies the choice of a relatively 
large association distance. 
In Figure~\ref{fuchiid} we show, for the 26 UC~HII regions associated with
an ISOGAL source, the distance between the radio and MIR positions as a 
function of the radio size. Radio sources with large diameter all have
larger association distances compared with the small diameter sources. 

\begin{figure}
\centerline{\psfig{figure=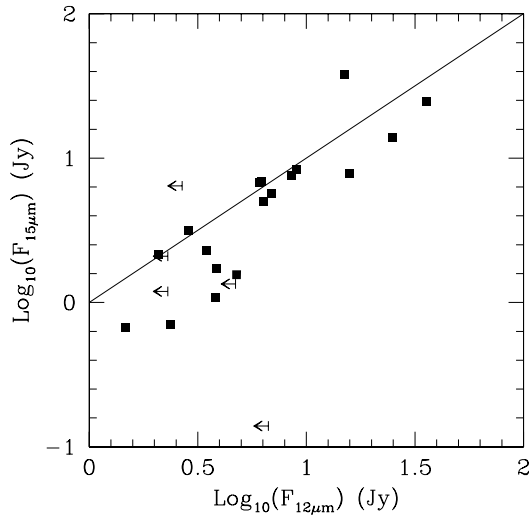,width=7.0cm}}
\caption[]{ISOGAL 15~$\mu$m flux density versus IRAS 12~$\mu$m flux density
for the 26 UC~HII associated with an ISOGAL-PSC source.}
\label{ff15f12}
\end{figure}

In Figure~\ref{ff15f12}, the 15~$\mu$m flux density of the 26 UC~HII are
compared with those at 12~$\mu$m from IRAS. There is a generally good
correlation between the two values, but, on the average, the 15~$\mu$m flux
densities are smaller than the 12~$\mu$m ones. This effect could be produced by
nearby point sources, unresolved in the IRAS beam, and/or by the extended
nature of the YSO, that could
be partly filtered out by the ISOGAL point source reconstruction routines.

%
%
%
%
%
%
%

\section{The selection criteria}
\label{sselc}

In Section~\ref{scrco} and Appendix~\ref{app_rnd}
we have shown that radio-IRAS selected UC~HII
and candidate UC~HII regions are reliably associated with a class of
ISOGAL sources that occupy a well defined region of the ([7]--[15],[15])
and (K--[7],[7]) colour magnitude diagrams.
This in itself does not guarantee that all ISOGAL sources with the same
colour-magnitude characteristics are YSOs. In Paper~II, also using the
results from Paper~I and from Testi et al.~(\cite{TFOPSCG97}), we showed that
all the three sources with [7]--[15]$\ge$1.8 and [15]$\le$4.5 (after correcting
for the offset discussed in Sect.~\ref{ig_phot_calib}) were bright YSOs.
In Paper~II we also used theoretical and empirical arguments to show that
the region of the ([7]--[15],[15]) colour-magnitude plane to the right
of the dashed line drawn in Figure~\ref{fcmd_ig_bwhz} is expected to
suffer by a very low contamination from Post-MS stars. Unfortunately,
if we wish to extract a list of YSO candidates from the entire
ISOGAL database, the criteria developed in Paper~II will have to be revised.
Compared to the $l$=$+$45$^\circ$ field discussed in our previous studies,
the ISOGAL fields close to the Galactic Centre are affected by a much larger
line-of-sight extinction, which has the effect of moving sources toward fainter
[15] magnitudes and slightly larger [7]--[15] colours (see the extinction vector
shown in Fig.~\ref{fcmd_ig_bwhz}b), with the net effect of
contaminating the region expected to be occupied by low flux density YSOs
with heavily reddened Post-MS stars. For this reason we decided to
restrict the search for YSOs only to the high flux density sources, i.e
sources with [15]$\le$4.5 mag (S$_{15}\ge$~330~mJy) and [7]--[15]$\ge1.8$.
Let us stress that such ISOGAL sources are much more luminous
than all YSOs studied by the ISOCAM Team group in nearby star forming
regions. For instance, a source with S$_{15}~\ge$~330~mJy,
even at the moderate distance for ISOGAL of 1.6~kpc, is more luminous than
all sources detected with ISOCAM in $\rho$ Oph
(d=160~pc) by Bontemps et al.~(\cite{Bont01}).

For the fields with DENIS observations
it is possible to use the near infrared magnitudes to add confidence
to the identification of the sources as probable YSOs. As shown in
Testi et al.~(\cite{TFOPSCG97}), such YSOs are expected to have K--[15]$>$9,
and, consequently, K--[7] larger than 4--7 (depending on the value of
[7]--[15], see Figure~\ref{fcmd_ig_bwhz}a). Since 
most of the K-band sources associated
with radio sources are probably fake associations (see Appendix~A), and 
even for reliable associations the interpretation of the near-infrared
data is complicated by the effect of extinction, we will not use the constraint
based on DENIS data to select the YSO candidates. In the final
YSO candidates table (Table~3) we will report 
whether the field has been observed at K and
if the source has a K--[7] colour exceeding 4 and [7]$<$6. The flag
S$_{\rm D}$ is set equal to ``1'' for sources satisfying these ISOGAL-DENIS 
confidence criteria, to ``0'' for sources not satisfying the criteria,
and to ``--1'' for sources within fields not covered by the DENIS survey.
Since bright, saturated stars cannot be extracted from the DENIS survey (see 
also Sect.~\ref{Denis_data}), we have checked the original DENIS images for
all the ISOGAL sources with [7]$<$6 without a DENIS counterpart in the 
catalogue. Indeed, for seven sources we either found a very bright, saturated
counterpart (in 4 cases) or an extended near infrared source (3 cases).
For these peculiar sources, in Table~3 instead of the DENIS
magnitudes we report an ``S'' (for saturated sources) or an ``E''
(for extended sources).

To check the reliability of the selection criteria and the contamination
from non-YSOs, we compare with the colours of sources in the high
galactic latitude fields and with the near-infrared
spectroscopic survey of Schultheis et al.~(\cite{Sea01}).
In high galactic latitude fields, we only find one object falling in the
candidate selection boxes. Since in these fields we do not expect to find a
significant number of luminous YSOs, but we still expect late
type Post-MS stars, they provide indirect
evidence that the number of evolved stars selected by our criteria
should be small.
This is also confirmed by recent near-infrared spectroscopic observations
of ISOGAL YSO candidates toward the inner regions of the Galaxy by Schultheis et
al.~\cite{Sea01}, who found strong contamination from Post-MS stars at faint
[15] magnitudes and bright K magnitude, but not within our revised selection
boxes shown in Figures~\ref{fcmd_ig_bwhz} and \ref{fdeniscmd}. However, these
spectroscopic observations did not address the reddest OH/IR stars which 
contaminate the YSO box, as discussed in Section~\ref{scont}.

The selection criteria have been defined to avoid as much as possible
the contamination from Post-MS stars. This necessary requirement has
important implications on the efficiency of the selection criteria.
As shown in Figure~\ref{fcmd_ig_bwhz} a non-negligible fraction of the
ISOGAL sources associated with radio-BWHZ identified YSOs to the left of
the [7]--[15]=1.8 line is rejected from our selection
criteria. As discussed in Sect~\ref{sigbwhz} a fraction of these are expected
to be false associations, especially those with blue colour and high [15]
magnitude. Considering as good only the ISOGAL-BWHZ matches with [15]$\le$4.5
and rejecting only the BWHZ source known to be associated with a PN, our
selection criteria reject $\sim$30\% of the {\it bona fide} radio selected YSOs.

The region to the right of the dashed line in Figure~\ref{fcmd_ig_bwhz}a and
below [15]=4.5, which was used in Paper~II to select low flux density YSOs in
the l=+45 fields, suffers from a contamination of reddened post-MS stars that
is too large to allow a reliable identification of YSOs towards the inner
regions of the Galactic Plane. For this reason we will not use it for the
purpose of this paper. With this choice we loose the low flux-density non-radio
emitting YSOs (corresponding to 90\% of the YSOs selected in Paper~II).

In Paper~II we have also included as possible YSOs those sources without
detection at 7~$\mu$m for which the lower limit to the [7]--[15] colour would
select them as YSO candidates. The vast majority of these sources (called
``candidate YSOs'' in Paper~II) had [15]$>$4.5. To increase the reliability of
the identification only sources with good detections at 7 and 15~$\mu$m are
considered in the present paper.

\section{The catalogue of bright YSO candidates}
\label{scat}

The selection criteria that we derived in the previous section imply that
we will restrict to the bright ([15]$\le$4.5) YSO candidates. This is
necessary to reduce to the minimum the number of spurious Post-MS sources
in the catalogue, but implies that we will limit the catalogue to high
mass YSO candidates plus a fraction of nearby intermediate mass YSOs.

The catalogue is presented in Table~3 (available only in electronic
form). All sources are detected at 7 and 15~$\mu$m and satisfy the
selection criteria based on the [15] magnitude and [7]--[15] colour defined
in Sect.~\ref{sselc}. For each source we also report the DENIS magnitudes
and the ISOGAL--DENIS association quality flag (a$_{ID}$), from the
ISOGAL catalogue
if the field has been observed by DENIS or ``--1'' otherwise
(see Omont et al. in preparation and Schuller et al. in preparation for details).
We also report a flag (S$_D$) that specifies whether the source satisfies the
([7],K-[7]) confidence criteria (see Sect.~\ref{sselc}), the flag is set
to ``1'' if the source satisfies the criteria, to ``0'' if it does not satisfy 
the criteria, or to ``--1'' if the field is not covered by the DENIS survey.

\subsection{Contamination from post-MS stars}
\label{scont}

As noted in section~\ref{sselc} and Appendix~\ref{app_rnd}, the colour cutoff 
[7]--[15]$\ge$1.8 has been introduced to avoid the contamination by the
overwhelmingly large population of post-MS stars in the YSO candidates 
catalogue. Clearly, the sharp cutoff used represent a trade-off between
low-contamination and high-efficiency in selecting YSO candidates.
Following the discussion in Glass et al.~(\cite{Gea99}) and Alard et
al.~(\cite{Aea01}), all the post-MS stars brighter than our [15] selection
threshold are expected to be high mass-losing AGB stars (with perhaps
a few post-AGB stars). The vast 
majority of AGB stars in our Galaxy are oxygen-rich, and practically
all of them are O-rich in the inner Galaxy where are located
most of the ISOGAL fields. Most of
the reddest and more extincted of these
are expected to be OH/IR stars (Pottasch~\cite{P93}).
Thus, to obtain an evaluation of the contamination from post-MS stars 
in our YSO candidates catalogue, we can compare our selection criteria
against OH/IR stars surveys. We can also use the longer wavelengths information
provided by the MSX mission (Price et al.~\cite{MSX}) for the 
brightest ISOGAL sources. 

\begin{figure}
\centerline{\psfig{figure=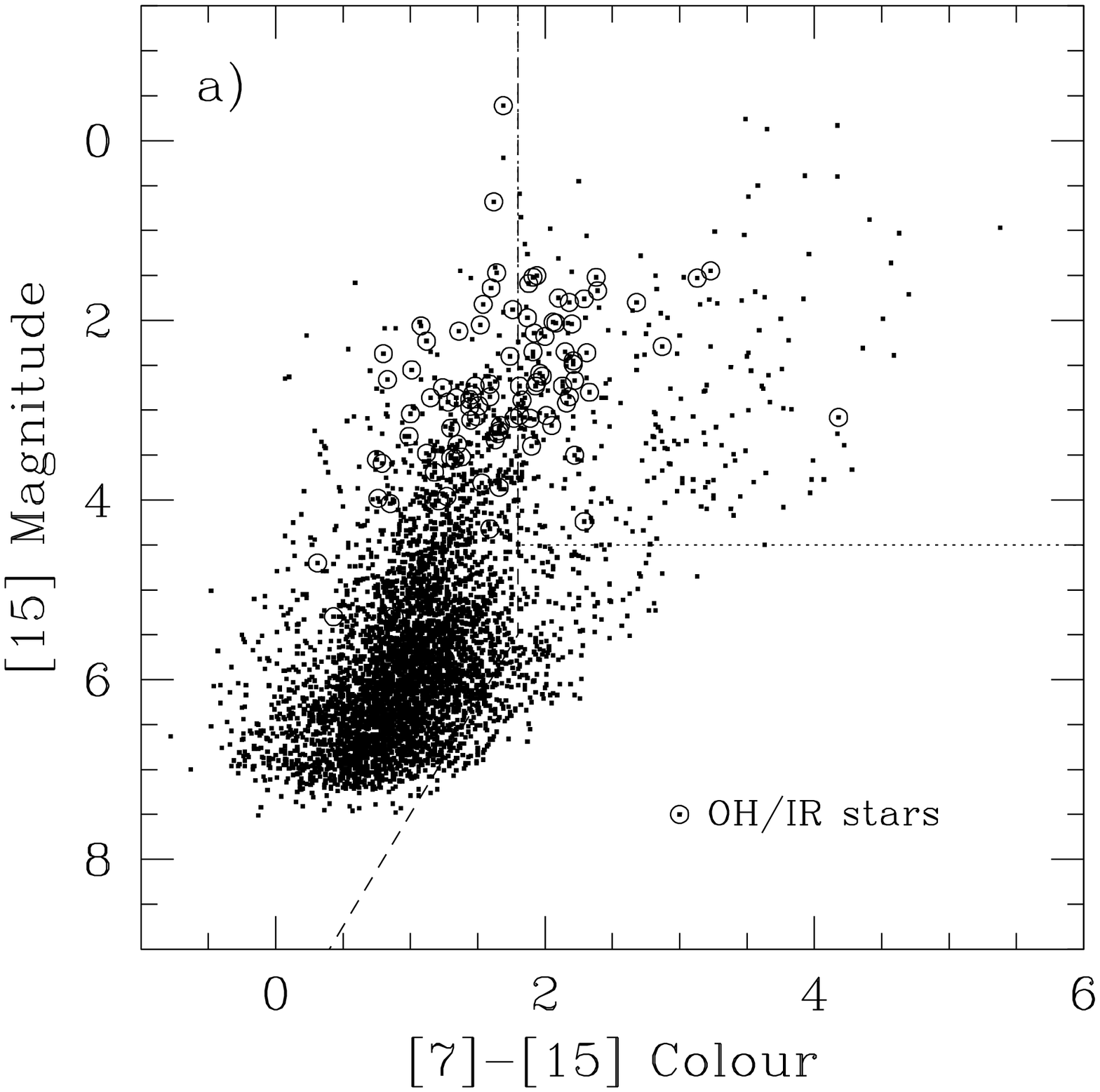,width=6.7cm}}
\centerline{\psfig{figure=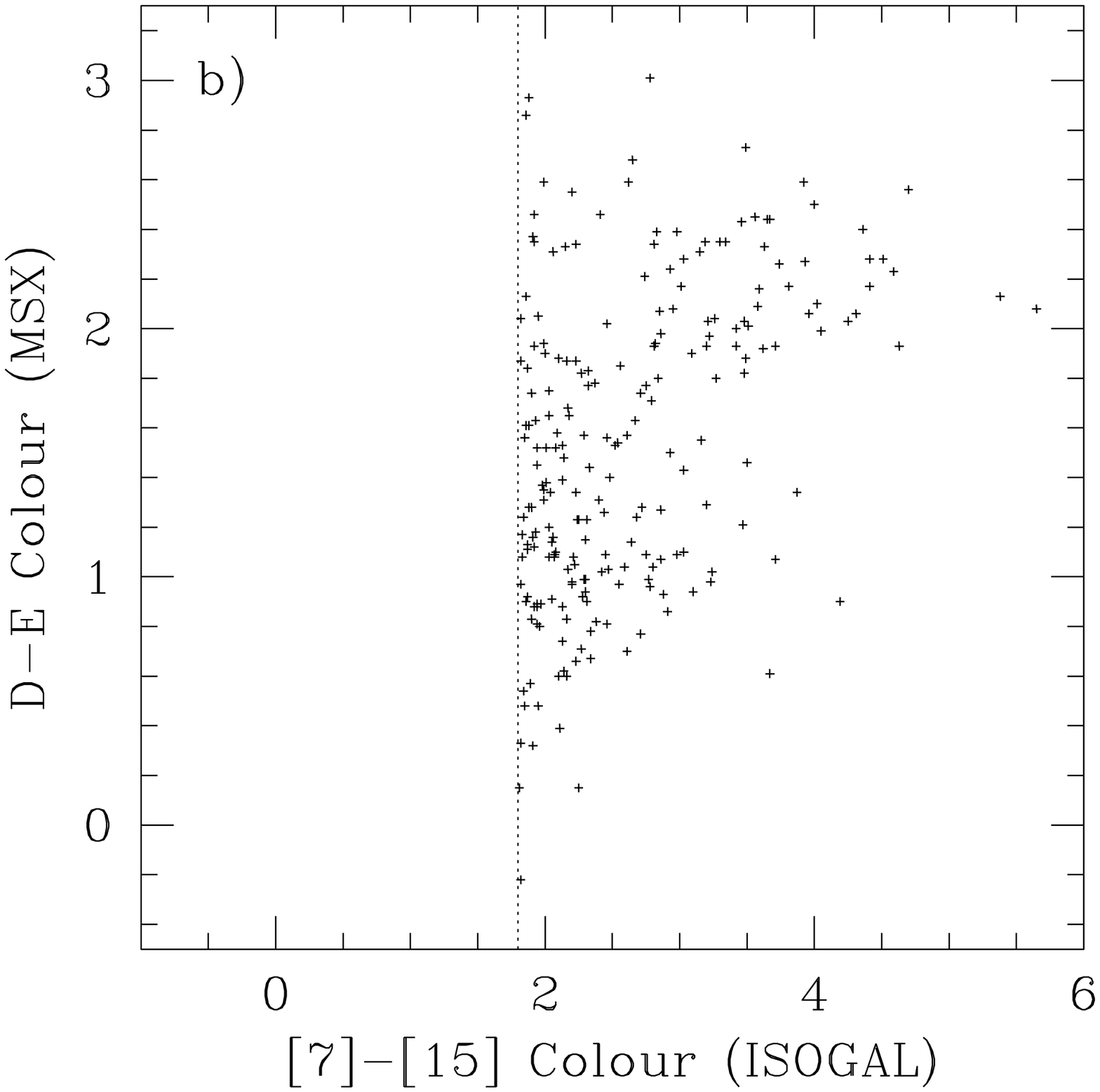,width=6.7cm}}
\centerline{\psfig{figure=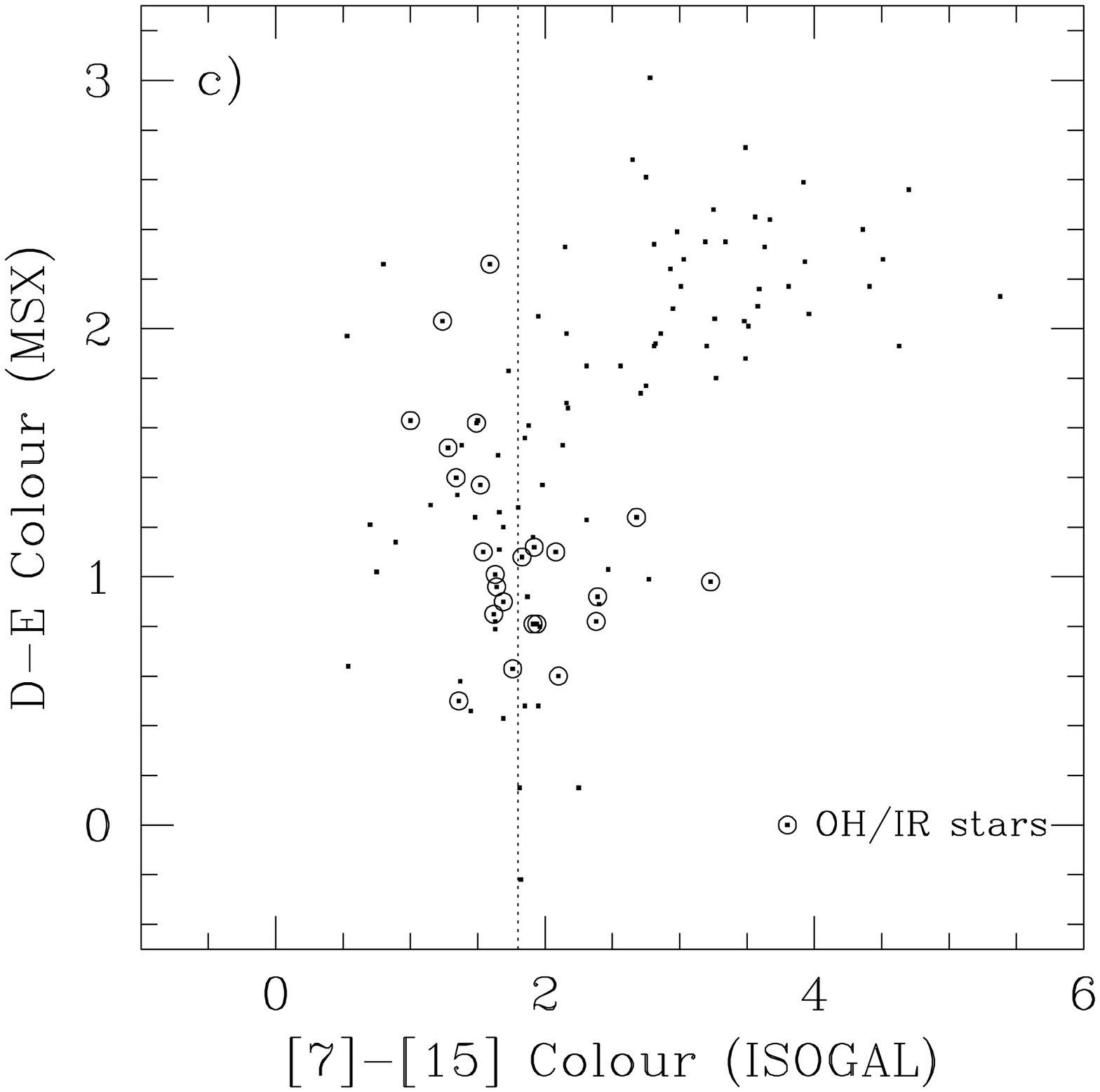,width=6.7cm}}
\caption{Contamination from post-MS stars of the YSO candidates catalogue.
{\bf a)} [15] vs. [7]--[15] colour magnitude diagram for ISOGAL
identified OH/IR stars (open circles) and for all ISOGAL sources
within the limits of the OH/IR surveys (filled squares).
The dotted and dashed lines are as in Figure~\ref{fcmd_ig_bwhz}.
{\bf b)} ISOGAL-MSX colour-colour diagram for all the ISOGAL
selected YSO candidates with MSX D and E counterpart;
{\bf c)} ISOGAL-MSX colour-colour diagram for all the ISOGAL sources
with MSX D and E counterpart within 
the limits of the OH/IR surveys (filled squares), the identified OH/IR stars
are indicated with open circles.
}
\label{fcont}
\end{figure}

Two catalogues of OH/IR stars in the central degree
around the Galactic Centre are given by Lindqvist et
al.~\cite{Lindq} and by Sjouwerman et al.~\cite{Sjouwer}.
In Figure~\ref{fcont}a, we show the positions of the
ISOGAL associated (with a 9'' search radius; Ortiz et al.~\cite{Oea02})
OH/IR stars in the [15] vs. [7]--[15] diagram
(open circles) superimposed on the distribution of all
ISOGAL sources within the limits of both OH/IR surveys.
As can be seen in this figure, nearly one half of the OH/IR
stars fall within our YSO candidate selection criteria.
Nevertheless, it should be noted that the variability of
this kind of stars and the non simultaneity of the 7~$\mu$m
and 15~$\mu$m observations can increase or reduce the observed
[7]--[15] colour (for instance, the variability can account
for one magnitude in [7]--[15] for the reddest OH/IR sources).
As opposite, only
17\% of the ISOGAL YSO candidates within the OH/IR stars survey area
are associated with an identified OH/IR star. Of course, the
OH surveys cannot be complete, and it is known that some post-MS sources
with infrared colours similar to OH/IR stars do not display OH emission,
so that this fraction is probably
underestimated, allowing that 20-25\% of the selected YSO candidates 
are probably very red post-MS stars.

Finally, additional information can be provided by the MSX
survey (Price et al.~\cite{MSX}), which has a sensitivity
about 3 magnitudes less deep than ISOGAL, but goes to longer
wavelength, with bands D and E, centred at 14.65 and 21.34
$\mu$m respectively. A detailed analysis of the MSX counterparts
of ISOGAL sources will be the object of a forthcoming paper,
but we can already show in Figures~\ref{fcont}b and~\ref{fcont}c the different
locations in the D--E vs. [7]--[15] colour-colour diagram of the
ISOGAL selected YSO candidates and the MSX and
ISOGAL associated OH/IR stars. Only a small
fraction (29\%) of the OH/IR stars have a D--E colour greater
than 1.3 (4 of these 7 sources have a bad quality detection
in one or the other band), and none of these stars has
[7]--[15] greater than 1.8. An association with MSX sources with a 10'' search
radius of the 715 ISOGAL selected YSO candidates results in
433 associations. Then, only 222 MSX sources have been detected
in the D and E bands, 129 of which (58\%) have a D--E colour
greater than 1.3. It seems from Figure~\ref{fcont}c that about 50\% of the
ISOGAL YSO candidates with D--E lesser than 1.3 are proved
post-MS stars; then, we find
that about 20\%, and maybe 30-35\%, of our sample should be
contaminated by post-MS stars.


This comparison shows that a lower contamination, in the direction of 
the Galactic Centre, could be obtained 
using a more conservative [7]--[15] cutoff, at the price of a reduced
efficiency in YSO candidates selection. For example, a cutoff of
[7]--[15]$\ge$2.5, would result in 
a lower efficiency in
recovering the {\it bona fide} radio selected YSOs (see sect~\ref{sselc} and
Fig.~\ref{fcmd_ig_bwhz}a), but the expected contamination in the
Galactic Centre region would drop to $\le$10\%.

\subsection{Global properties of the YSOs for the entire ISOGAL survey}
\label{sgp}

\begin{figure*}
\centerline{\psfig{figure=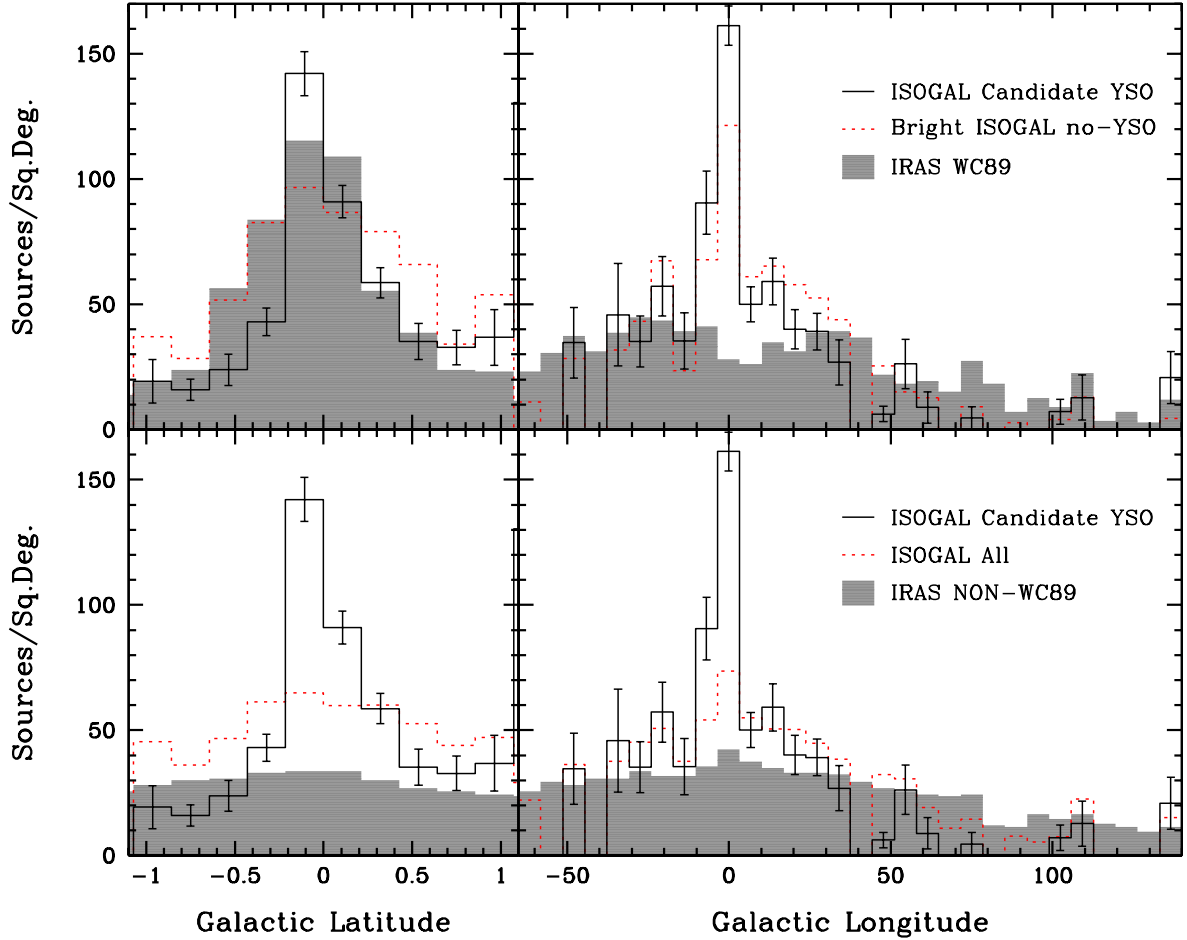,width=16cm}}
\caption[]{Galactic distribution of the bright YSO candidates compared with
other ISOGAL sources and IRAS-selected samples. All ISOGAL histograms have
been corrected for the uneven sampling of the Plane normalising each bin
by the area effectively observed. Bins for which the ISOGAL survey covered
less than 0.02~deg$^{2}$ have been set to null values. Top panels: the
distribution of YSO candidates is compared with the distribution of
bright ([15]$<$4.5) ISOGAL sources with [7]--[15]$<$1.8 (dotted line) and
with the distribution of IRAS sources satisfying the WC89
criteria for UC~HII regions (grey filled histogram);
all distributions are normalised to have the same integral as the YSO candidates
distribution. Bottom panels: the distribution of the YSO candidates is
compared with the distribution of all ISOGAL sources and of all IRAS sources
which do not satisfy the UC~HII criteria. In both cases the IRAS sources are
extracted only in the region: $-$65$^\circ\le l\le$140$^\circ$,
$-$3$^\circ\le b\le$3$^\circ$.}
\label{fysolb}
\end{figure*}

Using our selection criteria, we extracted a total of 715 YSO candidates
from the ISOGAL-PSC, corresponding to $\sim$1\% of all entries 
in the PSC or $\sim$2\% of the sources with good detections at 7 and
15~$\mu$m.
From our sample of 715 YSO candidates, 525 (73\%) have a DENIS K
counterparts, and 78 (15\%) have K--15$>$8. Since most K non-detections
should have K--15 greater than 6-8 (depending on the exact limiting magnitude
for DENIS-ISOGAL association),
we can estimate that 30-35\% of the YSO candidates
have K--15$>$8. 
In Figure~\ref{fysolb} we show the galactic distribution of the sample
of YSO candidates (solid histograms with error bars),
selected with the criteria discussed above.
The histograms have been corrected for the uneven sampling of the
ISOGAL survey (see Sect.~\ref{sigcat}), i.e. the counts in each
latitude or longitude bin have been corrected for the area effectively
observed in that bin by ISOGAL. The error bars are poissonian errors
on the source counts in each bin corrected for the same area factor.
The area covered in each bin ranges from $\sim$2.8 to 0.02~sq.deg., bins with
smaller coverage are set to null values.

In the top panels of Figure~\ref{fysolb}, the histograms for the YSO candidates
are compared with the distribution of bright ([15]$\le$4.5) ISOGAL sources
with [7]--[15]$<$1.8 (i.e. rejected from our selection criteria), corrected for
the observed area in the same way as the YSOs, and with the distribution of
IRAS sources satisfying
the WC89 colour criteria for UC~HII. The latitude histogram shows that the
selected YSO candidates have a distribution highly peaked on the Galactic Plane
(as expected for young sources), and very similar to the distribution of
IRAS-selected UC~HII regions (grey full histogram). The bright ISOGAL sources
rejected by our colour selection criterion (dotted histogram) show a broader
latitude distribution, confirming our expectation of a substantial
contamination from evolved stars.  The longitude distributions show a clear
peak of the YSO candidates toward the Galactic Centre, a similar peak is
missing in the IRAS-selected UC~HII because of confusion problems in the
densest regions of the Galaxy.  The incompleteness of the IRAS sample of UC~HII
was also noted by BWHZ.

The bottom panels of Figure~\ref{fysolb} show the comparison between the
distribution of ISOGAL YSO candidates and the distributions of all
the ISOGAL sources (including sources detected only at 7~$\mu$m,
dotted histogram) and of all IRAS-PSC sources on the
Galactic Plane that do not satisfy the WC89
criteria for UC~HII regions. The comparison shows that while the
YSO candidates are highly concentrated on the Galactic Plane and towards the
Galactic Centre, the other two samples show a much broader distribution,
as expected for evolved objects.


As a word of caution, we wish to stress that the ISOGAL database  suffer 
of potential biases induced by the observational strategy. As already 
noted above, the survey deliberately avoids bright IRAS sources,
and bright star forming regions, in a non-uniform way: close to the 
Galactic Centre, narrow band filters were used, and the IRAS flux 
limit criterion to avoid saturation was relaxed. Moreover, given the patchy
coverage of the Plane out of the immediate surroundings of the Galactic Centre,
small features in the l and b distributions of Figure~\ref{fysolb}, such as
the slight asymmetry of the b distribution or the small peaks in the l
distribution, are not reliable. Some of these features may be caused 
by the sparse sampling of the ISOGAL survey, for instance, the small peak 
at l$\sim$15$^\circ$ may be related to the M16 star forming region, which 
was deliberately covered by one of the ISOGAL rasters.


\section{Conclusions}

In this work we have extended and brought to its conclusion the problem,
already approached in Paper~II, of identifying YSO candidates from the 
much larger population of MIR sources (predominantly Post-MS stars) found during the 
ISOGAL mapping at 7 and 15~$\mu$m of selected regions of the Galactic Plane
with ISOCAM. 

The selection criteria proposed in Paper~II, and verified there from the 
coincidence of ISOGAL selected bright YSOs with thermal radio continuum 
sources, have now been tested against a much larger sample of radio-IRAS 
identified YSOs in the Galactic Plane, by cross-correlating ISOGAL sources 
detected at 7 and 15~$\mu$m with the list of UC~HII regions from the VLA 5~GHz
Galactic Plane survey of BWHZ.
A statistical simulation has been implemented to establish the reliability 
of the radio-ISOGAL cross-correlation as a function of various parameters.
The results confirm that ISOGAL sources with an associated radio-loud YSO occupy
a well defined region of the ([7]--[15],[15]) colour-magnitude diagram and that
this region is relatively well separated from that occupied by the much larger
population of Post-MS stars. A similar segregation of the radio identified
YSOs detected also at K-band occurs in the (K--[7],[7]) colour-magnitude diagram.
However, the near infrared criteria cannot be used for the entire 
ISOGAL catalogue since the DENIS observations only cover the fields located at
$\delta\le2^\circ$. Additionally, most of the K-band sources associated
with radio sources are probably fake associations (see sect.~\ref{sselc} and
Appendix~\ref{app_rnd}), and 
even for reliable associations the interpretation of the near-infrared
data is complicated by the effect of extinction.
Therefore we retained the near infrared information only as
``confidence criteria'' for the sources for which there are available data.

With the aim of providing a more reliable list of YSO candidates
throughout the Galactic
Plane observations of ISOGAL, following the indications of the comparison with
the large sample of radio identified YSOs, we have revised the selection
criteria adopted in Paper~II, which have been restricted to [15]$\le$4.5 and
[7]--[15]$\ge1.8$, with the additional confidence criteria of [7]$\le$6 and
K--[7]$\ge$4. This choice is motivated by the fact that for larger [15]
magnitudes, toward the inner regions of the Galactic Plane, the contamination
from reddened Post-MS stars becomes too large to obtain a reasonable list of
candidates. The application of those criteria to the entire ISOGAL catalogue
(only fields observed both at 7 and 15~$\mu$m were considered) has produced a
list of 715 YSO candidates, which represent $\sim$2\% of the total number of
ISOGAL point sources detected at both wavelengths.  All the three bright,
point-like YSOs identified in Paper~II are reselected using our new criteria
(and the revised ISOGAL-PSC).  The galactic distribution of the selected
sources is strongly peaked on the Galactic Plane, as expected for very young
sources, and show a clear peak close to the Galactic Centre. Our results
confirm previous suggestions by BWHZ that the IRAS sample of massive YSOs
selected using the WC89 criteria is severely limited by confusion in the inner
regions of the Galaxy.

Since the separation between YSOs and Post-MS stars in the colour-magnitude
plane uses by necessity a sharp function, we expect a contamination of non-YSOs
in our list of candidates of at least 20\%, as well as we expect that we may
have missed some YSOs.  Using our selection criteria we recover $\sim$70\% of
the radio-identified YSOs. The limitation to bright YSOs ([15]$\le$4.5) is
particularly needed in the ISOGAL fields close to the Galactic Centre, where
large line-of-sight extinction is expected, and the contamination from reddened
Post-MS stars at high [15] magnitudes is particularly high.  This restriction
could be released in less extincted regions, such as the l=+45$^{\circ}$ fields
discussed in Paper~II. In high-extinction fields, a lower contamination fraction
could be obtained by increasing the [7]--[15] cutoff limit, at the price of a
lower YSO selection efficiency. 

\begin{acknowledgements}
We thank the survey preparation, observations planning, data reduction and
calibration ISOGAL team for the excellent, dedicated work culminated in the
production of the Point Source Catalogue.
Support from CNR-ASI grant 1/R/27/00 to the Osservatorio
Astrofisico di Arcetri is gratefully acknowledged.
\end{acknowledgements}

\appendix
\section{Reliability of the ISOGAL-BWHZ samples association}
\label{app_rnd}

In order to quantitatively estimate the reliability of the
association between ISOGAL point sources and BWHZ radio continuum
sources, we performed a detailed test of
ISOGAL sources associated with random samples.
The random samples were designed to mimic the
galactic distribution of the BWHZ sample.
For this purpose we limited the BWHZ sample to the formal survey
boundaries ($-10^\circ\le l< 40^\circ$ and $|b|< 0.4^\circ$),
which implies a reduction of the BWHZ sample to a total of 1171 sources.
The galactic distribution has been approximated by a constant
function of the longitude and a combination of a constant plus a
Gaussian in latitude. The parameters of the Gaussian and constant
distribution were chosen to fit the observed sources distribution.
Using a long periodicity pseudo-random number generator (Press et
al.~\cite{Pea92}) we generated 10000 samples of 1171 sources distributed
within the boundaries of the BWHZ formal survey.
The random samples were then
correlated with the ISOGAL database following the same criteria
as for the real BWHZ sample and using an association radius of 15\arcsec.

\begin{figure*}
\centerline{\psfig{figure=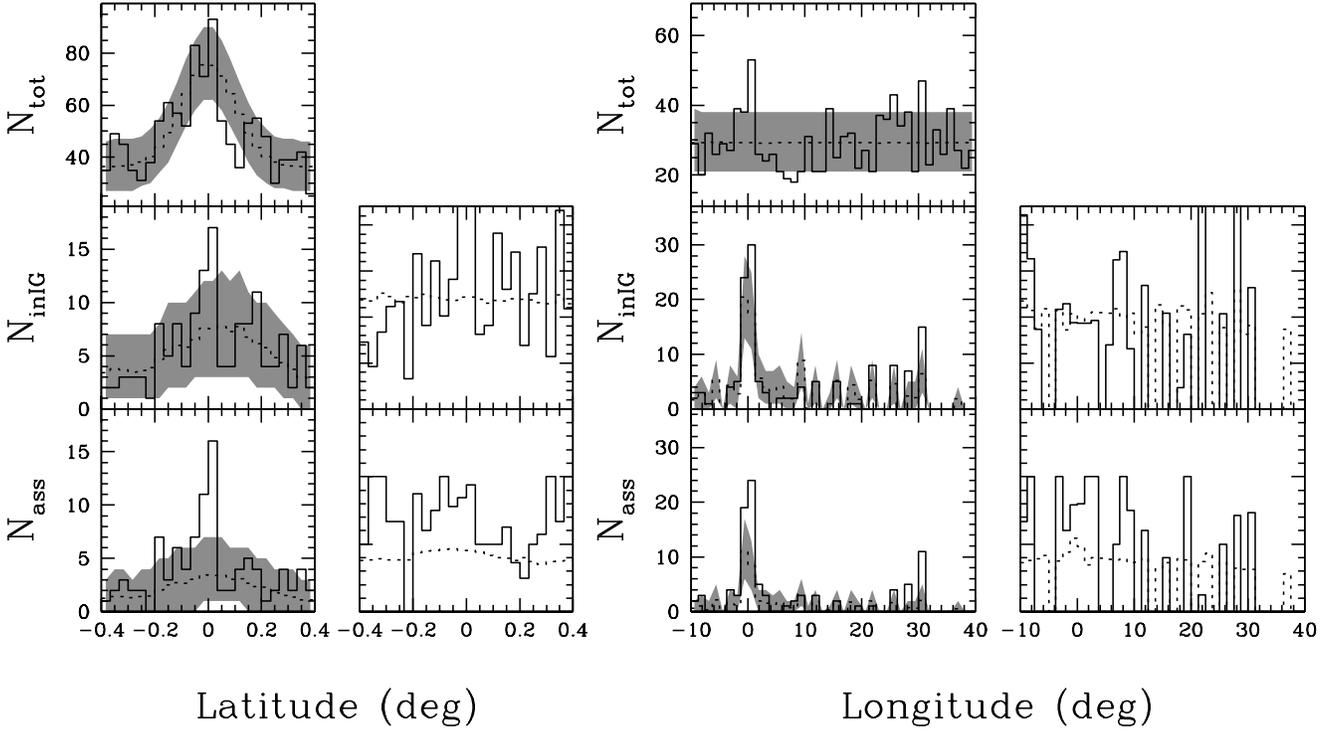,width=17.6cm}}
\caption[]{Galactic latitude and longitude distributions for the
BWHZ and the random samples. In all panels, the solid histogram shows the
distribution for the BWHZ sample, the dotted histogram is the average of all
the random samples, the gray area shows, for each bin, the region containing
90\% of the random realizations. The top panels show the distributions
for the entire samples, the middle panels those of the sources within the
ISOGAL fields, and the bottom panels those of the sources with an associated
ISOGAL source.
The unlabelled panels to the right of the distributions in
the middle and bottom rows show the rates of sources within ISOGAL fields
corrected for the area covered by the ISOGAL field in each latitude and
longitude bin and the association rates respectively.}
\label{lbplot}
\end{figure*}

In Figure~\ref{lbplot} we show the galactic distribution of the
BWHZ sample (solid histograms) the average distribution of the
random samples (dotted histograms) and, for each bin, the shaded region
represent the range covered by 90\% of the random realizations.
The top panels show the distributions for all sources within the
formal survey boundaries, the middle
panels for the sources within the ISOGAL fields, and the bottom ones
the distributions for the sources associated with ISOGAL sources.
The histograms illustrate that the simple functions used to describe
the galactic distribution of the BWHZ sample represent a reasonably accurate
approximation. Nevertheless, one should note two features in the BWHZ
distributions that are not represented in the random samples: a five peaks
modulation in the latitude distribution and a small, narrow peak at
l$\sim$0$^{\circ}$ in the longitude distribution. The five peaks in the latitude
distributions at b$=$0.0$^{\circ}$,$\pm$0.16$^{\circ}$,$\pm$0.32$^{\circ}$
correspond to the five centre
positions of the BWHZ pointing rasters, where the sensitivity of their
observations is higher, while the peak at the Galactic Centre is probably a
real feature of the radio sources distribution. For the purpose of the
random simulations we decided to ignore these two features, as a proper
reproduction of the latitude and longitude distributions
would require a proper modelling of the BWHZ sensitivity throughout
their observed region (see their Figure~1), the modelling of the radio
sources luminosity function and their longitude distribution. The complexity
and uncertainties of such detailed modelling go far beyond what is needed for
our purpose. A possible effect of the neglected features combined with the
uneven distribution of the ISOGAL fields is a moderately higher number of
radio sources within ISOGAL fields in the real BWHZ sample with respect to the
random samples, due to the crowding of ISOGAL fields near
the Galactic Centre position (see Figure~\ref{fields}).
The histograms of the galactic distribution of sources within ISOGAL fields
and of sources associated with ISOGAL sources show a number of features
which are probably related to the peculiar distribution of radio (and random)
sources and the uneven sampling of the Plane by ISOGAL observations.
To check whether the number of sources within ISOGAL fields and the number of
sources associated with ISOGAL sources are a function of galactic position
one has to properly correct for the number of radio (and random) sources per
coordinate bin and for the area covered by the ISOGAL survey in each
particular bin. These corrected ``rates'' of sources within ISOGAL fields
and of sources associated with an ISOGAL source are also shown in
Figure~\ref{lbplot}
(the histograms have null values in longitude bins
not covered by ISOGAL fields). These histograms show that, within the noise
and once appropriately corrected, the rates are constant
as a function of galactic position. A clear feature is that the rate of
BWHZ sources associated with ISOGAL sources is higher than the average rate
for the random samples.

\begin{figure}
\centerline{\psfig{figure=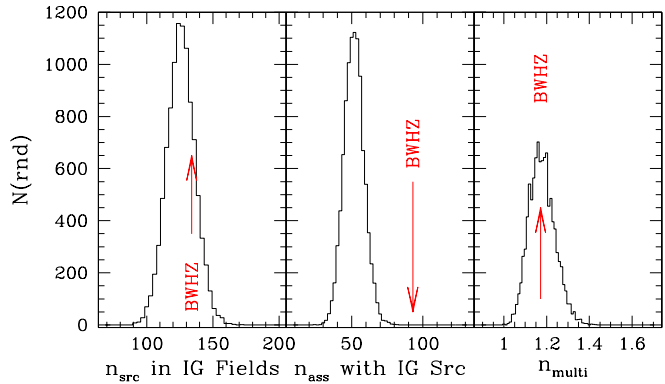,width=8.8cm}}
\caption[]{Distributions of the number of sources within ISOGAL fields (left),
number of sources associated with at least one ISOGAL source (middle), and
number of ISOGAL sources within 15\arcsec\ per source (right), for each
of the 10000 random realizations. The values for the BWHZ sample are indicated
with the arrows.
}
\label{rnd_count}
\end{figure}

In Figure~\ref{rnd_count} we present, for the 10000 random samples, the
distributions of the number of sources within ISOGAL fields (left),
number of sources associated with at least one ISOGAL source (middle), and
number of ISOGAL sources which can be associated with each BWHZ (or random)
source using an association radius of 15\arcsec\ (right). The corresponding
values for the BWHZ sample are also shown.
As expected, due to the effect of the uneven ISOGAL sampling and the small
peak at $l\sim$0$^\circ$ in the BWHZ sample, the number of BWHZ sources
within ISOGAL fields is marginally higher than for the random samples.
Quantitatively, the number of BWHZ sources within ISOGAL fields is 134,
7\% more than the median and barely within the third quartile of
the random realizations. The higher rate of ISOGAL sources associated with
BWHZ with respect to the random sources shown in Figure~\ref{lbplot} and
noted above is also clearly visible in the second panel of
Figure~\ref{rnd_count}, the association rate with the real sources is
almost double than with the random sources. Finally, the third panel
of Figure~\ref{rnd_count} shows that, within the selected association radius
(15\arcsec), as much as 15\% of the radio sources can be associated with
more than one ISOGAL source. The fraction of multiple associations in the BWHZ
sample is consistent with the random simulations suggesting that multiple
associations are not real ones, or that there is no crowding of ISOGAL sources
close to the radio positions on the association radius scale.

\begin{figure}
\centerline{\psfig{figure=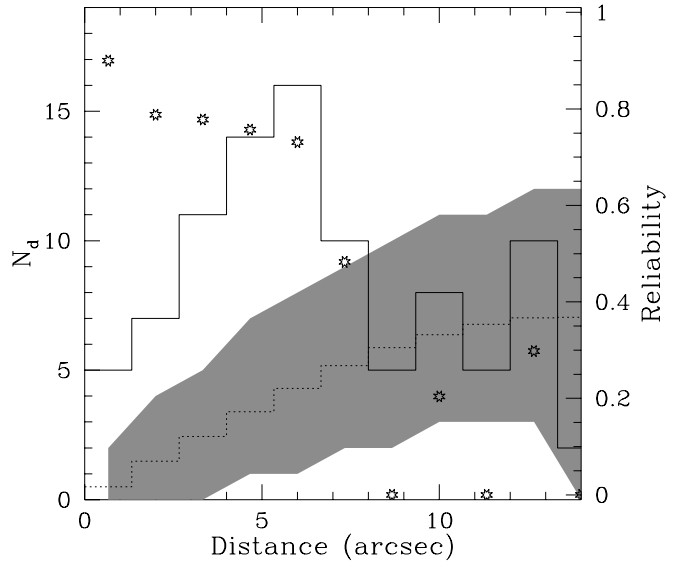,width=8.8cm}}
\caption[]{Distribution of the distances between radio source and associated
ISOGAL source, solid histogram for the BWHZ sample, dotted histogram for
the average of the random samples, the shaded area shows, for each bin, the
range covered by 90\% of the random realizations (ordinate axis on the left).
The stars show the reliability figure for each bin, computed as described in the
main text (ordinate axis on the right)}
\label{dplot}
\end{figure}
The distribution of the distances between the radio source and associated
ISOGAL source is reported in Figure~\ref{dplot}. The BWHZ sample is shown as
a solid histogram, the average of the random samples as a dotted histogram and,
for each bin, the grey region shows the location of 90\% of the random
realizations. The figure shows, as expected, that the number of random associations
increases steadily with the association distance. The BWHZ sample
has most of the associations for distances less than 7 arcsec, where, in each
bin, it clearly displays an excess of associations with respect to the
random samples. For larger distances the association rate of the BWHZ sample is
essentially indistinguishable with the expectations from the random samples.
To quantify the excess of associations in the BWHZ sample with respect to the
random expectations, for each distance bin we defined a reliability figure
as the difference between the number of BWHZ sources and the average of the
random realizations normalised to the number of BWHZ sources in that bin
(the figure is set to zero if the random associations are more than the BWHZ
ones). With this definition, for each bin the reliability figure gives
the expected fraction of ``real'' (i.e. non-random) associations in the
BWHZ sample. The reliability is plotted in Figure~\ref{dplot} as open stars,
and shows that for distances less than approximately 7\arcsec\ the contamination
from random associations is always expected to be less than 25\%, above
8\arcsec\ the number of associations in the BWHZ sample is indistinguishable
from those of the random realizations. These results are consistent with
simple association probability calculations based on the average densities
of ISOGAL sources reported in the ancillary tables of the ISOGAL public
catalogue (Omont et al. in preparation, Schuller et al. in preparation).

As discussed in Sect.~\ref{sigbwhz}, the above conclusions consider only one
aspect and may be misleading if taken alone: the chosen association distance
should also take into account the physics of the sources.
\begin{figure*}
\centerline{\psfig{figure=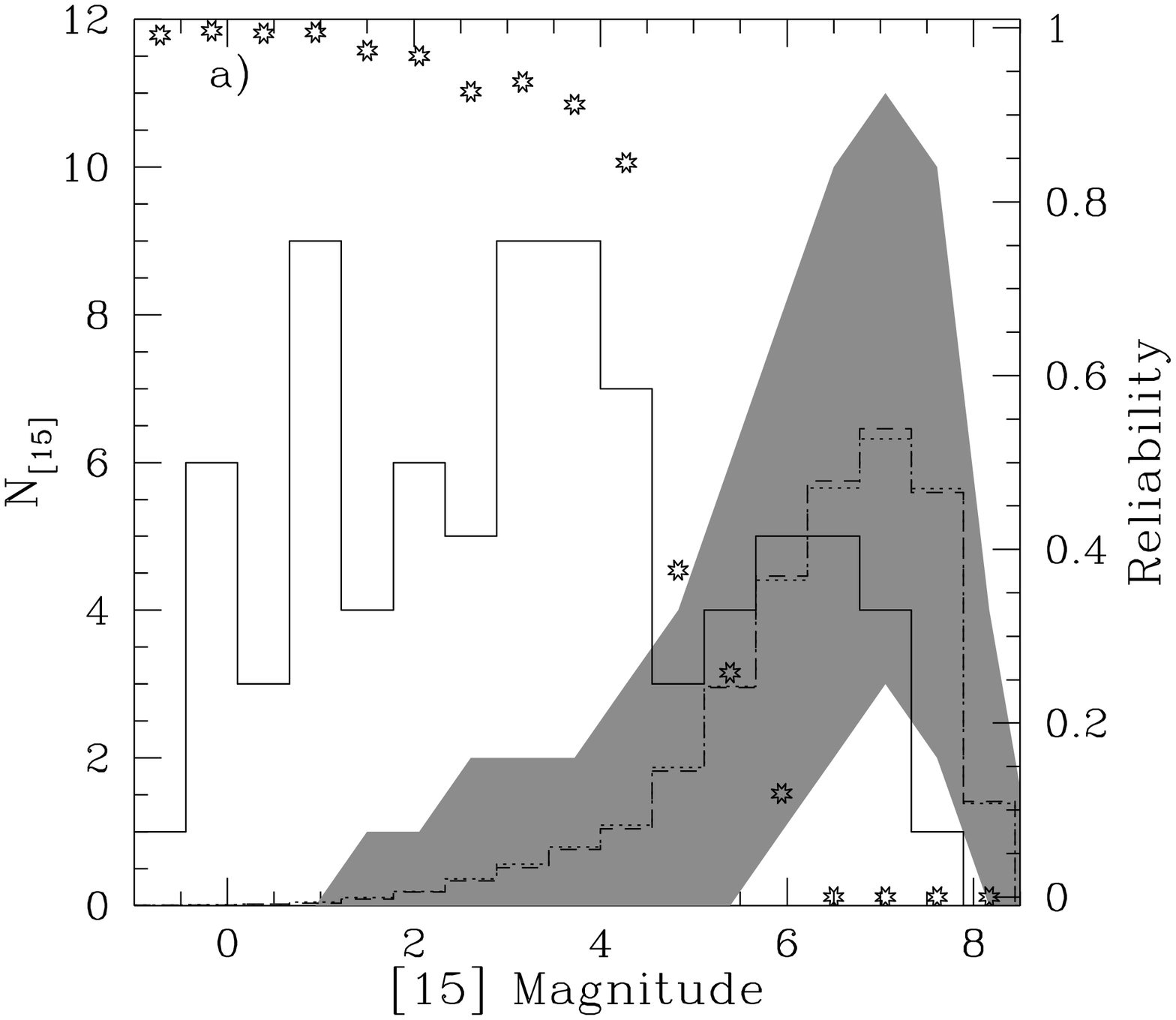,height=5.6cm}
            \psfig{figure=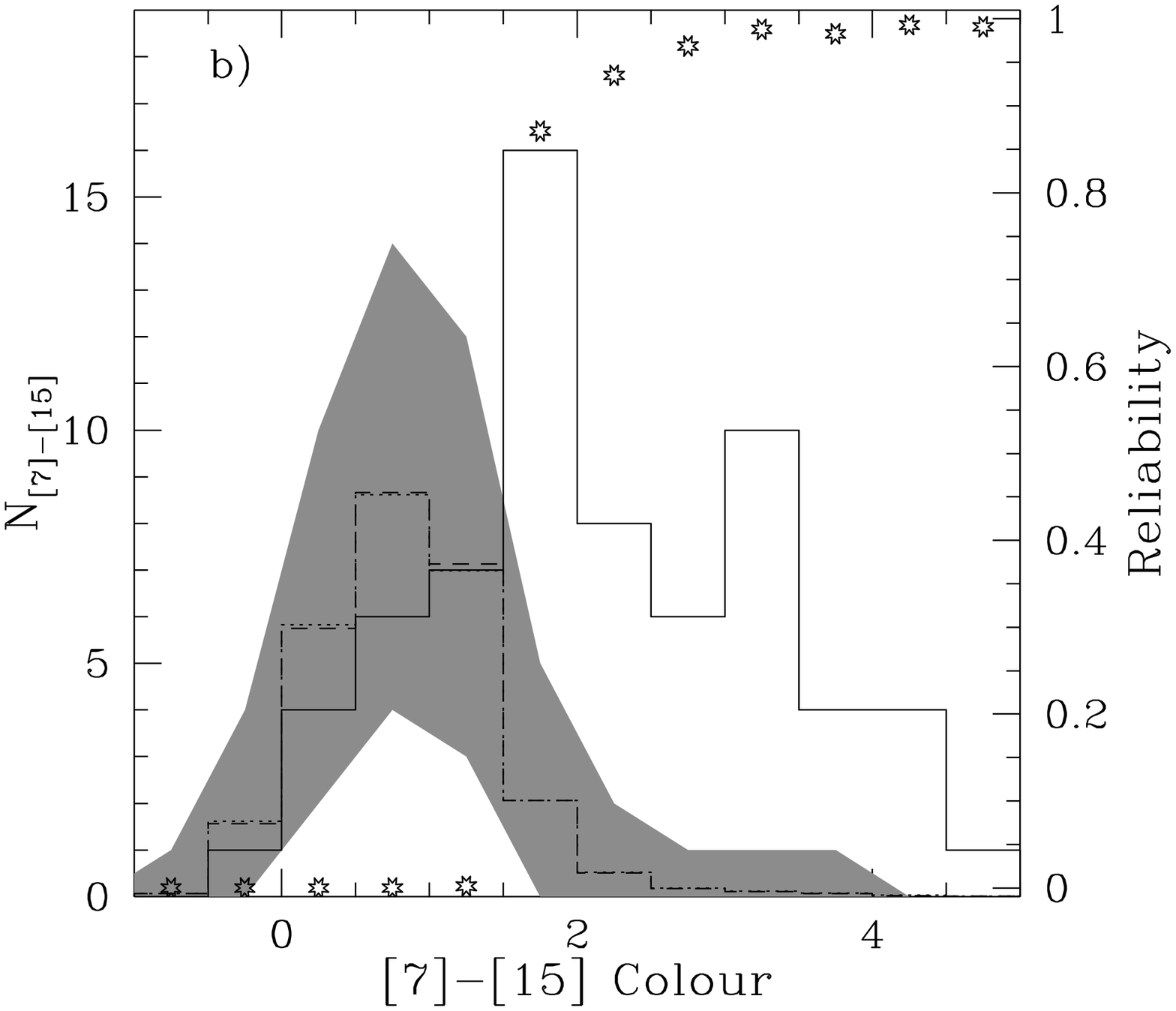,height=5.6cm}
            \psfig{figure=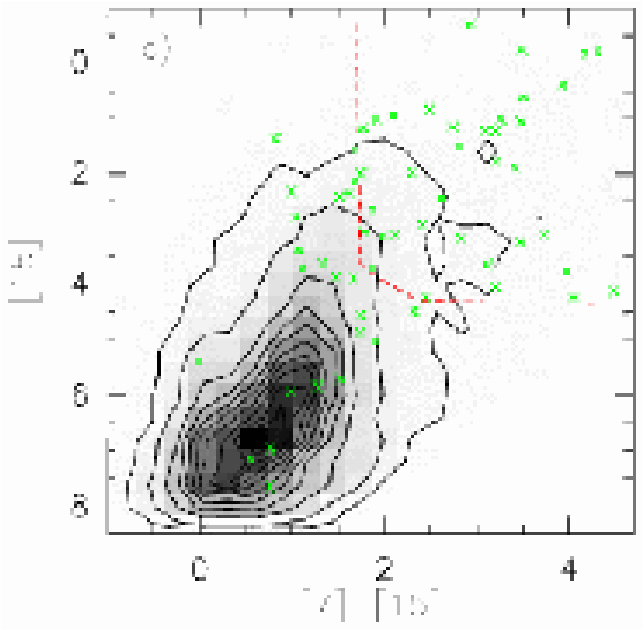,height=5.6cm,angle=-90}}
\caption[]{a) Distribution of the 15~$\mu$m magnitudes for the associated
ISOGAL sources, solid histogram for the BWHZ sample, dotted histogram for
the average of the random samples, the shaded area shows, for each bin, the
range covered by 90\% of the random realizations (ordinate axis on the left).
The dashed histogram shows the distribution of the 15~$\mu$m magnitudes of
all the ISOGAL sources within the formal BWHZ area, normalised to the same
total number of sources as the average distribution of the random samples.
The stars show the reliability figure for each bin, computed as described in the
main text (ordinate axis on the right). b) Same plot but for the [7]$-$[15]
colour index. c) Average colour-magnitude diagram for the sources associated with
the random samples (grey scale and contour plots); the sources associated
with the BWHZ sample are shown as crosses. Contour levels are 1\%, 5\% to 95\%
every 10\% of the peak value. The dashed line indicates the box where the
product of the reliability figures derived from the magnitudes and colours
histograms exceeds 80\%.}
\label{f715plots}
\end{figure*}
\begin{figure*}
\centerline{\psfig{figure=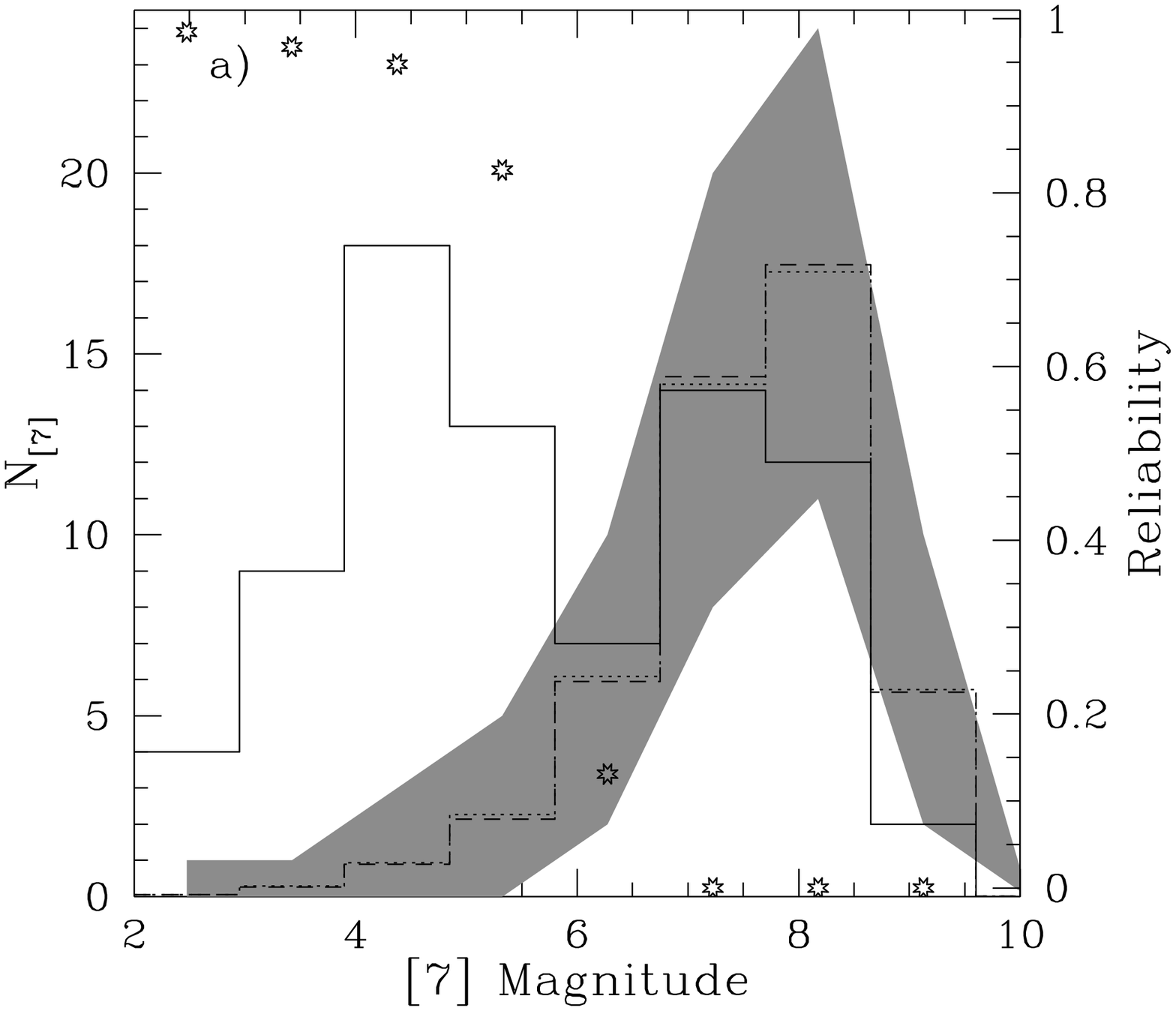,height=5.6cm}
            \psfig{figure=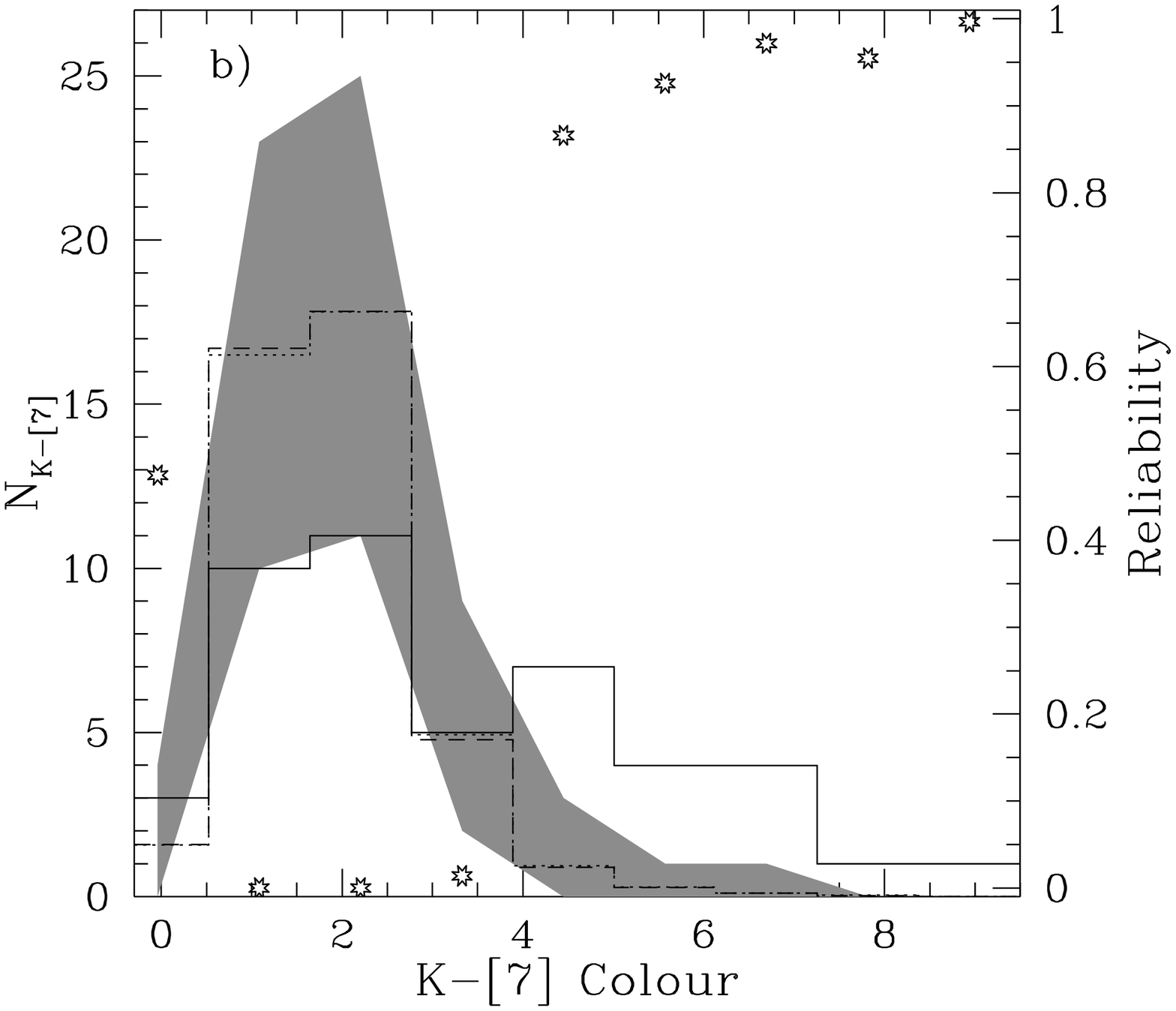,height=5.6cm}
            \psfig{figure=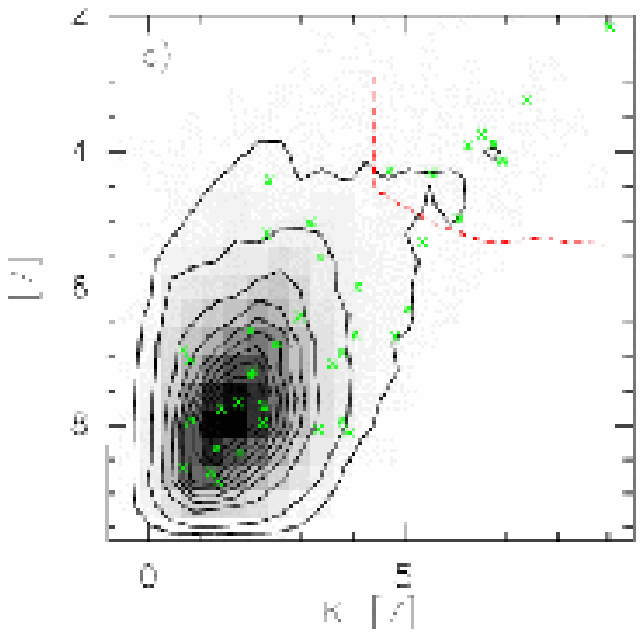,height=5.6cm,angle=-90}}
\caption[]{As Figure~\ref{f715plots} but for the [7] magnitude, K--[7]
colour, and (K--[7],[7]) colour-magnitude diagram.}
\label{fk7plots}
\end{figure*}
A better test can be performed on the brightnesses and colours of the ISOGAL
sources associated with the BWHZ and the random samples.
In Figures~\ref{f715plots}a
and~\ref{f715plots}b we show the distributions of the ISOGAL 15~$\mu$m
magnitudes and [7]$-$[15] colours. The definition of the
symbols is the same as for Figure~\ref{dplot}. In this figures we also
show, as a dashed line, the histogram of the magnitudes and colours of all the
ISOGAL sources within the formal BWHZ survey region, normalised
to the total number of sources in the average of the random samples. As
expected, the dotted and dashed histograms are nearly identical.
In summary,
the BWHZ sources are mostly associated with bright and red sources,
which are a minority of the ISOGAL sources, while the random samples
are mostly associated with the much larger population of
faint bluer sources. As a consequence,
the reliability figures show a sharp transition in both plots, suggesting a
very high probability of non-random associations for sources brighter
than [15]$\sim$5 and redder than [7]$-$[15]$\sim$1.5 and a comparably low
probability for fainter and bluer sources.
The same result is shown in a different graphical form in
Figure~\ref{f715plots}c,
where the average colour-magnitude diagram produced from the
sources associated with the random simulations is compared with the
position of the sources associated with the BWHZ sample. Most of the ISOGAL
sources associated with the BWHZ sample fall in a region of the diagram
where one would expect very few random associations, confirming that most of the
associations between ISOGAL sources and BWHZ radio sources are not chance
coincidences.

In Figure~\ref{fk7plots} we show similar figures for the [7] magnitude,
K--[7] colour, and (K--[7],[7]) colour magnitude diagram. While the [7] magnitude
shows a similar trend as the [15] or [7]--[15] colour, with most of the ISOGAL
sources associated with the BWHZ sources
populating the low-probability region of the plot, the situation is
less clear for the K--[7] colour. Most of the sources {\it with a good K
magnitude in the catalogue} populate the region expected for random
associations, and a few sources populate the region of the plot with
high reliability of being non-random associations. As discussed in
Sect.~\ref{sigbwhz}, we believe that this is a consequence of the
nature of the infrared sources reliably associated with BWHZ sources:
these are characterised by very red colours, and in many case are not
detectable at K by the DENIS survey. Consequently, most of the ISOGAL sources
{with a good K magnitude} plotted in Figure~\ref{fk7plots} are probably
fake ISOGAL-BWHZ
associations, as confirmed by the high value of the [7] magnitude and the
large association distance (see Sect.~\ref{sigbwhz}).

Combining all the results of our simulations, we conclude that the ISOGAL
sources with the highest probability of being reliably associated with a
radio source are those with bright [15] magnitudes and large [7]--[15]
colours. The (K--[7],[7]) colour-magnitude diagram can be used as an
additional indication of a good association if the particular ISOGAL field
has been observed by DENIS and if the source is not detected or is very faint
at K. To offer a more quantitative estimate for these reliability criteria,
in Figures~\ref{f715plots}c and \ref{fk7plots}c we mark with a dashed line
the region of the colour-magnitude diagrams where the product of the
reliability figures for both magnitude and colour exceeds 80\%. ISOGAL sources
associated with BWHZ sources above and to the right of the lines have
the highest reliability of being true associations, and should be used
to derive the infrared properties of radio sources.




\begin{thebibliography}{}
\bibitem[2001]{Aea01}
  Alard C., et al. 2001, ApJ 552, 289
\bibitem[1993]{AWTB93}
  Andr\'e P., Ward-Thompson D., Barsony M., 1993, ApJ 406, 122
\bibitem[1999]{AWTB99}
 Andr\`e P., Ward-Thompson D., Barsony M., 1999,  in Protostars and
  Planets IV, eds. S. Mannings, A. Boss, and S. Russell
  (Tucson: Univ. Arizona Press), p.~59
\bibitem[2000]{BAPABW}
 Bacmann A., Andr\`e P., Puget J.-L., Abergel A., Bontemps S., Ward-Thompson D.,
2000, A\&A 361, 555
\bibitem[1994]{bec94}
  Becker R.H., White R.L., Helfand D.J., Zoonematkermani S. 1994, ApJS 91, 347
\bibitem[1990]{BWMHZ90}
  Becker R.H., White R.L., McLean B.J., Helfand D. J.,
Zoonematkermani S., 1990, ApJ 358, 485
\bibitem[1992]{BCILNS92}
 Berrilli F., Corciulo G., Ingrosso G., Lorenzetti D., Nisini B., Strafella
F., 1992, ApJ 398, 254
\bibitem[2001]{Bont01}
 Bontemps S., Andr\'e P., Kaas A.A. et al., 2001, A\&A 372, 173
\bibitem[1998]{Cea98}
 Carey S. J., Clark F. O., Egan M. P.  et al. 1998, ApJ 508, 721
\bibitem[1991]{C91}
 Churchwell E., 1991, in The Physics of Star Formation and Early Stellar
Evolution, C.J. Lada and N.D. Kylafis eds., Kluwer Academic Publishers, 221
\bibitem[1990]{CWW90}
 Churchwell E., Wolfire M. G., Wood D.O.S., 1990, ApJ 354, 247
\bibitem[2000]{CNK00}
 Comer\'on F., Neuh\"auser R., Kaas A.A. 2000, A\&A 359, 269
\bibitem[1988]{Eea88}
 Eder J., Lewis B.M., Terzian Y., 1988, ApJS 66, 183
\bibitem[1994]{DENIS}
 Epchtein N., de Batz B., Copet E., et al., 1994, Ap\&SS, 217, 3
\bibitem[1999]{DENIS2}
 Epchtein N., Deul E., Derriere S., et al., 1999, A\&A, 349, 236
\bibitem[1998]{FCHHLR98}
 Faison M., Churchwell E., Hofner P., Hackwell J., Lynch D.K., Russel R.W.,
1998, ApJ 500, 280
\bibitem[1998]{Fea98}
  Feldt M., Stecklum B., Henning Th., Hayward T.L., Lehmann Th., Klein R.
    1998, A\&A 339, 759
\bibitem[1999]{Fea99}
  Feldt M., Stecklum B., Henning Th., Launhardt R., Hayward T.L.
    1999, A\&A 346, 243
\bibitem[1979]{felli}
  Felli M.,  1979, in Stars and Stellar Systems, Bengt E. Westerlund Ed.
\bibitem[2000]{FCTOS00}
  Felli M., Comoretto G., Testi L., Omont A., Schuller F. 2000, A\&A 362, 199
  (Paper~II)
\bibitem[1991]{Fea91}
  Fomalont, E.B., Windhorst, R.A., Kristian, J.A., Kellerman, K.I. 1991,
    AJ 102, 1258
\bibitem[1999]{Gea99}
  Glass I.S., Ganesh S., Alard C., Blommaert J.A.D.L., Gilmore G., et al. 
    1999, MNRAS 308, 127
\bibitem[1992]{HZBW92}
  Helfand D. J., Zoonematkermani S., Becker R.H.,  White, R.L.,
1992, ApJS 80, 211
\bibitem[1992]{HSVK92}
 Hillenbrand L.A., Strom S.E., Vrba F.J., Keene J., 1992, ApJ 397, 613
\bibitem[1997]{IE97}
 Ivezi\'c Z., Elitzur M., 1997, MNRAS 287, 799
\bibitem[2001]{kk}
 Kim K.T., Koo B.C., 2001, ApJ 549, 979
\bibitem[1994]{KCW}
 Kurtz S., Churchwell E., Wood D.O.S., 1994, ApJS 91, 659
\bibitem[1987]{LADA87}
 Lada C.J., 1987, in ``Star Forming Regions'', IAU 115, p. 1
\bibitem[1999]{Lada99}
 Lada C.J., 1999, The Origin of Stars and Planetary System, C J. Lada and
   N.D. Kylafis eds., NATO Science Series vol. C-540 (Dordrecht: Kluwer), p.~143
\bibitem[1984]{LW84}
 Lada C.J. \& Wilking B.A., 1984, ApJ 287, 610
\bibitem[1992]{Lindq}
Lindqvist M., Winnberg A., Habing H.J., Matthews H.E.,
1992, A\&ASS 92, 43
\bibitem[1999]{MIVE}
 Miroshnichenko A., Ivezsi\'c Z., Vinkovic D., Elitzur M., 1999, ApJ 520, L115
\bibitem[1998]{MTBCP}
 Molinari S., Testi L., Brand J., Cesaroni R., Palla F., 1998,
ApJ 505, L39.
\bibitem[1999]{Natta99}
 Natta A., 1999, Infrared space astronomy, to-day and to-morrow, F. Casoli,
F. David and J. Lequeux eds., EDP-Sciences, Springer-Verlag
\bibitem[2001]{NT01}
 Natta A. \& Testi L. 2001, A\&A 376, L22
\bibitem[1996]{Nordh96}
 Nordh L., Olofsson G., Abergel A. et al., 1996, A\&A 315, L185
\bibitem[1999]{Olofsson99}
Olofsson G., Huldtgren M., Kaas A.A. et al., 1999, A\&A 350, 883
\bibitem[2002]{Oea02}
 Ortiz R., Blommaert J.A.D.L., Copet E., Ganesh S., Habing H.J.,
 Messineo M., Omont A., Schultheis M., Schuller F. 2002, A\&A 388, 279
\bibitem[2000]{Pea00}
  Persi P., Marenzi A.R., Olofsson G., Kaas A.A., Nordh L., et al.
   2000, A\&A 357, 219
\bibitem[1973]{pan}
 Panagia N., 1973, AJ 78, 929
\bibitem[1996]{Pea96}
  P\'erault M., Omont A., Simon G., S\'eguin P., Ojha D., et al.
    1996, A\&A 315, L165
\bibitem[1997]{PSL97}
 Pezzuto S., Strafella F., Lorenzetti D., 1997, ApJ 485, 290
\bibitem[1993]{P93}
 Pottasch S.R. 1993, in {\it Infrared Astronomy}, A.~Manpaso, M.~Prieto and
   F.~S\'anchez Eds. (Cambridge University Press: Cambridge, UK), p.~63
\bibitem[1988]{Pea88}
 Pottasch S.R., Bignell C., Olling R., Zijlstra A.A., 1988, A\&A 205, 248
\bibitem[1992]{Pea92}
 Press W.H., Teukolsky S.A., Vetterling W.T., Flannery B.P., 1992, ``{\it
 Numerical Recipes in C}'', Second Edition, Cambridge University Press
\bibitem[2001]{MSX}
 Price S.D., Egan M.P., Carey S.J., Mizuno D.R., Kuchar T.A., 2001,
 AJ 121, 2819
\bibitem[1985]{RL85}
 Rieke G.H. \& Lebofsky M.J. 1985, ApJ 288, 618
\bibitem[1980]{RR80}
 Rowan-Robinson M., 1980, ApJS 44, 403
\bibitem[2002]{Sea01}
 Schultheis M., Parthasarathy M., Omont A., Cohen M., Ganesh S.,
 Sevre F., Simon G. 2002, A\&A 386, 899
\bibitem[1976]{SK76}
 Scoville N.Z., Kwan J., 1976, ApJ 206, 718
\bibitem[1998]{Sjouwer}
 Sjouwerman L.O., van Langevelde H.J., Winnberg A., Habing H.J.,
  1998, A\&ASS 128, 35
\bibitem[1999]{USNO}
 Stone R. C., Pier J. R., Monet D. G., 1999, AJ, 118, 2488
\bibitem[1998]{TFPR98}
  Testi L., Felli M., Persi P., Roth M., 1998, A\&A 329, 233
\bibitem[1997]{TFOPSCG97}
 Testi L., Felli M., P\'erault M., S\'eguin P., Omont A., Comoretto G.,
  Gilmore G., 1997, A\&A 318, L13
\bibitem[1999]{TFT99}
 Testi L., Felli M., Taylor G. 1999, A\&AS 138, 71
  (Paper~I)
\bibitem[2002]{Tea02}
 Testi L., Natta A., Oliva E., D'Antona F., Comer\'on F., Baffa C.,
    Comoretto G., Gennari S. 2002, ApJ 571, L155
\bibitem[1995]{Tea95}
 Testi L., Olmi L., Hunt L., Tofani G., Felli M., Goldsmith P.
    1995, A\&A 302, 249
\bibitem[1997]{Wea97}
 Watson A.M., Coil A.L., Shepherd D.S., Hofner P., Churchwell E.
   1997, ApJ 487, 818
\bibitem[1999]{Wea99}
 Walsh A.J., Burton M.G., Hyland A.R., Robinson G.
   1999, MNRAS 309, 905
\bibitem[2001]{wbbn}
 Walsh A.J., Bertoldi F., Burton M.G., Nikola T., 2001, MNRAS 326, 36
\bibitem[1991]{WBH91}
  White R.L., Becker R.H., Helfand, D. J., 1991,  ApJ 371, 148
\bibitem[1989]{WC89}
  Wood, D.O.S., Churchwell, E. 1989, ApJ 340, 265, (WC89)
\bibitem[1989]{WC89s}
  Wood, D.O.S., Churchwell, E. 1989, ApJS 69,831
\bibitem[1990]{ZHB90}
  Zoonematkermani S., Helfand D.J., Becker R.H., White R.L., Perley R.A.,
  1990, ApJS 74, 181
\end{thebibliography}
\end{document}